\documentclass[10pt,a4paper]{article}

\usepackage{jheppub}
\usepackage[T1]{fontenc} 
\usepackage{epsfig,epsf}
\usepackage{amsmath}
\usepackage{amsthm}
\usepackage{amsfonts}
\usepackage{amssymb}
\usepackage{caption}
\usepackage{dsfont}
\usepackage{mathrsfs}
\usepackage{mathtools}
\usepackage{epstopdf}
\usepackage{multirow}

\usepackage{rotating}
\usepackage{marvosym}

\usepackage{soul}
\usepackage{xcolor}
\usepackage{breqn}
\usepackage{subcaption}
\usepackage{array}   
\newcolumntype{C}{>{$}c<{$}}

\usepackage{slashed}
\usepackage{booktabs}
\makeatletter
\renewenvironment{pmatrix}[1][1]{%
  \left(\mkern-1mu\renewcommand{\arraystretch}{#1}\begin{array}{*{\c@MaxMatrixCols}{c}}%
}{%
  \end{array}\mkern-1mu\right)%
}
\makeatother

\def\lgen{\mathcal{L}^{\scriptstyle\text{gen}}}

\title{
{\textsc
Three-loop anomalous dimensions of leading-twist operators in the Gross-Neveu-Yukawa model}
}

\author[]{Mrigankamauli Chakraborty}
\author[]{and Leonid A. Shumilov}

\affiliation[]{
   II. Institut f\"ur Theoretische Physik, Universit\"at Hamburg,
   D-22761 Hamburg, Germany}

\emailAdd{mrigankamauli.chakraborty@desy.de}

\emailAdd{leonid.shumilov@desy.de}

\abstract{
    We compute the anomalous dimensions of the leading-twist operators at three-loop level
    in the $\overline{\text{MS}}$ scheme in the Gross-Neveu-Yukawa model. Calculations are done
    for a number of low-spin operators using computer algebra systems, and the analytical form is
    reconstructed by solving a set of Diophantine equations. Calculations are optimized using the emergent supersymmetry of the corresponding generalized theory. For the obtained results, we perform a comprehensive analysis at the critical point, including a comparison with the $1/N$ expansion, a check of the generalized Gribov-Lipatov reciprocity, and an analysis of the behavior next to the rightmost singularity. The latter allows us to extract the values of Regge intercepts for all the leading-twist trajectories in the form of an $\epsilon$ expansion.  
    }

\preprint{PUBDB-2026-02879}

\begin{document}
\maketitle

\newpage
\section{Introduction}
The operator product expansion (OPE) is a powerful tool used in phenomenological and theoretical studies of quantum field theories (QFTs). In quantum chromodynamics (QCD), the Wilson OPE~\cite{Wilson:1969zs} allows one to separate perturbative and non-perturbative contributions to the hadronic cross-sections. In conformal field theories (CFTs), the conformal OPE serves as a crucial conceptual object~\cite{Polyakov:1970xd} and leads to the existence of important technical tools, such as the conformal bootstrap~\cite{Rattazzi:2008pe}. In both cases, the use of OPE requires knowledge of corrections to the canonical dimensions of the involved operators --- their anomalous dimensions. Due to the special structure of the OPE, the leading contribution is given by the operators of leading twist, where twist is the difference between the scaling dimension and the spin. 

As one of the most technically complex cases, QCD sets the state of the art for the calculations of leading-twist anomalous dimensions. The 
long-standing three-loop result~\cite{Moch:2004pa, Vogt:2004mw} was only recently pushed to four-loop order in the flavor non-singlet case~\cite{Gehrmann:2026qbl}. Despite the advances in computer algebra systems, these calculations remain challenging, and become hardly feasible for the operators of large spin. One possible way to bypass this complication is to ``bootstrap'' the analytic form of the result based on the data for the (finite) number of low-spin operators. This method was successfully implemented, see e.g.~\cite{Velizhanin:2012nm, Moch:2014sna, Kniehl:2025ttz} for the use in QCD and~\cite{Velizhanin:2013vla, Marboe:2016igj, Kniehl:2021ysp} in the $\mathcal{N} = 4$ SYM. Using the structure of anomalous dimensions as functions of spin, one creates an ansatz and then fixes coefficients by solving the set of Diophantine equations using the LLL algorithm~\cite{Lenstra:1982eee}. In view of this procedure, specific properties of anomalous dimensions, such as Gribov-Lipatov reciprocity~\cite{Basso:2006nk,Dokshitzer:2006nm}, play an important role.    

In this paper, we calculate the leading-twist anomalous dimensions in the Gross-Neveu-Yukawa (GNY) model~\cite{Zinn-Justin:1991ksq} at three-loop order using the above-mentioned bootstrap approach. This model serves as a UV completion of the two-dimensional Gross-Neveu model~\cite{Gross:1974jv}. Below four dimensions it possesses an IR-stable critical point (see e.g.~\cite{Zerf:2017zqi}), which is believed to describe the phase transition in graphene~\cite{Herbut:2006cs}. The renormalization of the GNY model was recently done at the five-loop level~\cite{Gracey:2025aoj}, and anomalous dimensions of leading-twist operators were previously known with two-loop accuracy~\cite{Manashov:2025kgf}. From a broader perspective, GNY at the critical point serves as an example of a fermionic CFT, and leading-twist anomalous dimensions in this regard can be seen as additional data for the numerical conformal bootstrap, see e.g.~\cite{Erramilli:2022kgp}. Another remarkable property of the model is that, when continued to the fractional number of flavors, $N_f = 1/4$, it becomes supersymmetric~\cite{Grover:2012bm, Fei:2016sgs}.      

Besides the interest from the condensed matter perspective, we use the GNY model as a convenient testing ground for the possible optimization of the calculation procedure of leading-twist anomalous dimensions, as well as the deeper studies of their analytic properties as functions of spin. In particular, in~\cite{Chakraborty:2026lna} the optimization method based on emergent supersymmetry was suggested. The GNY model in this approach is seen as a specific case of a broader theory described by the so-called generalized Lagrangian, including also the Nambu-Jona-Lasinio-Yukawa (NJLY) model~\cite{Fei:2016sgs} and the $\mathcal{N} = 1$ Wess-Zumino model~\cite{Wess:1974tw}. The emergent supersymmetry of the generalized Lagrangian allows one to connect the Green's functions involving different operators in the flavor singlet leading-twist sector, which notably reduces the complexity of calculations. Note that in the case of gauge theories, a similar generalized Lagrangian connecting $\mathcal{N} = 1, 2, 4$ SYM theories also exists~\cite{Chakraborty:2026wie}. In connection to the analytic structure of anomalous dimensions, in this paper we focus on their behavior around the rightmost singularity, following~\cite{Manashov:2025kgf}. This behavior at the critical point turns out to be constrained by the structure of analytical continuation of CFT in spin~\cite{Caron-Huot:2017vep, Kravchuk:2018htv}. In particular, the singularity is associated with level-crossing with the shadow trajectory, see~\cite{Caron-Huot:2022eqs} for details. Resolving this degeneracy allows us to perform a resummation. This, in turn, leads to the prediction of the singular behavior of the higher-order terms, valuable in view of the LLL-based analytic reconstruction. Additionally, we use the resummed anomalous dimensions to extract the positions of the Regge intercepts~\cite{Costa:2012cb}. The rightmost singularity in the GNY model shares common features with the rightmost singularity in the flavor non-singlet anomalous dimension in QCD, and the described resummation procedure explains the existence of the so-called generalized double-logarithmic equation (DLE)~\cite{Velizhanin:2014dia}.

The paper is organized as follows. In Section~\ref{sec:background} we give the definition of the GNY model and set up the problem of calculating the leading-twist anomalous  dimensions. Additionally, we discuss the critical properties of this model and its relation to the generalized theory possessing supersymmetry. In Section~\ref{sec:results} we describe the computational framework together with original details on the optimization of standard techniques using emergent SUSY. We also provide the actual results of the calculations in this section. In Section~\ref{sec:checks} we discuss the expected properties of the results and verify perfect agreement. The set of properties includes the relation to the conserved currents, complementary data from the framework of the $1/N$ expansion, the generalization of Gribov-Lipatov reciprocity, and the validity of emergent SUSY constraints. Section~\ref{sec:small-spin} is devoted to the behavior of the results around the rightmost singularity. We discuss its connection to the conformal Chew-Frautschi plot, perform the resummation procedure, and extract valuable information from the resummed results. In Section~\ref{sec:sum} we summarize. Appendices are reserved for the technical details. 

\section{Background\label{sec:background}}
\subsection{The Gross-Neveu-Yukawa model}
We consider the model in the $d = 4 -2\epsilon$ dimensional Euclidean space described by an action
\begin{equation}
\label{eq:gny-action-def}
    S = \int d^dx \left(\bar{q}\slashed{\partial}q  + \dfrac{1}{2}(\partial \sigma)^2 + g\bar{q}\sigma q + \dfrac{\xi}{4!}\sigma^4\right),
\end{equation}
where $q_i(x)$ ($i = 1, \ldots, N$) is an $N$-component fermion field and $\sigma(x)$ is a real scalar field. This model is multiplicatively renormalizable 
\begin{align}
    q \mapsto Z_qq, && \sigma \mapsto Z_{\sigma}\sigma, && g \mapsto Z_gg, && \xi \mapsto Z_{\xi}\xi,
\end{align} 
and the renormalized action reads
\begin{equation}
  S_{R} = \int d^dx \left(Z_1\bar{q}\slashed{\partial}q  + \dfrac{Z_2}{2}(\partial \sigma)^2 + Z_3\mu^{\epsilon}g\bar{q}\sigma q + Z_4\mu^{2\epsilon}\dfrac{\xi}{4!}\sigma^4\right),
\end{equation}
with
\begin{align}
  Z_1 = Z_q^2, && Z_2 = Z_\sigma^2, && Z_3 = Z_gZ_q^2Z_\sigma, && Z_4 = Z_{\sigma}^4Z_{\xi},
\end{align}
and $\mu$ is the renormalization scale. It is convenient to define a rescaling of the couplings and introduce
\begin{align}
  u = \dfrac{g^2}{(4\pi)^2}, && \lambda = \dfrac{\xi}{(4\pi)^2}.
\end{align}
In the paper we work in the $\overline{\text{MS}}$ scheme, so the renormalization factors, $Z$, take the form 
\begin{equation}
    Z(u, \lambda; \epsilon) = 1 + \sum_{k = 1}^{\infty}\epsilon^{-k}Z_{k}(u, \lambda),
\end{equation}
where $Z_{k}(u, \lambda)$ are $\epsilon$-independent constants calculated as a series in $u$ and $\lambda$.  
RG functions in the model~\eqref{eq:gny-action-def} are known to five-loop accuracy~\cite{Gracey:2025aoj}. Here we provide the relevant expressions up to three loops
\allowdisplaybreaks{\begin{subequations}
\label{eq:rg-functions}
\begin{align}
\label{eq:beta-u}
\beta_u(u, \lambda) \equiv \dfrac{du}{d\ln \mu} &= -2\epsilon u + (4N + 6)u^2 - \left(24N + \dfrac{9}{2}\right)u^3 - 4u^2\lambda + \dfrac{u\lambda^2}{6} \nonumber \\ &\quad  + \left(28N^2 + \left(\dfrac{67}{4} + 108\zeta_3\right)N - \dfrac{697}{8} + 114\zeta_3\right)u^4 \nonumber \\ &\quad + (30N + 42)u^3\lambda - \left(\dfrac{5}{4}N - \dfrac{91}{24}\right)u^2\lambda^2 - \dfrac{u\lambda^3}{8} + \ldots,  \\
\label{eq:beta-lambda}
\beta_\lambda(u, \lambda) \equiv \dfrac{d\lambda}{d\ln \mu} &= -2\epsilon\lambda - 48Nu^2 + 8Nu\lambda + 3\lambda^2 + 384Nu^3 - \dfrac{17}{3}\lambda^3 + 28Nu^2\lambda - 12Nu\lambda^2 \nonumber \\ &\quad - \left(3768N^2 - \left(30 - 2304\zeta_3\right)N\right)u^4 + \left(868N^2 - \left(\dfrac{4395}{2} + 936\zeta_3\right)N\right)u^3\lambda, \nonumber \\ &\quad  -\left(36N^2 - \left(\dfrac{361}{2} + 324\zeta_3\right)N\right)u^2\lambda^2 + \dfrac{43}{2}Nu\lambda^3 + \left(\dfrac{145}{8} + 12\zeta_3\right)\lambda^4 + \ldots,\\
\label{eq:gamma-q}
\gamma_{q}(u, \lambda) \equiv \dfrac{d \ln Z_{q}}{d\ln \mu} &= \dfrac{u}{2} -  \left(\dfrac{3}{2}N + \dfrac{1}{8}\right)u^2 - \left(\dfrac{3N^2}{2} - \dfrac{47}{8}N + \dfrac{15}{32} - \dfrac{3}{2}\zeta_3\right)u^3 + u^2\lambda - \dfrac{11}{96}u\lambda^2 + \ldots, \\ 
\label{eq:gamma-sigma}
\gamma_{\sigma}(u, \lambda) \equiv \dfrac{d \ln Z_{\sigma}}{d\ln \mu} &= 2Nu - 5Nu^2 + \dfrac{\lambda^2}{12} + \left(25N^2 + \left(\dfrac{21}{8} + 6\zeta_3\right)N\right)u^3 \\ \nonumber &\quad + 5Nu^2\lambda - \dfrac{5}{8}Nu\lambda^2 - \dfrac{\lambda^3}{16} + \ldots.
\end{align}
\end{subequations}}
In what follows we also need the anomalous dimension of the operator $\sigma^2(x)$, which reads~\footnote{Note that here we mean the anomalous dimension of the local composite operator $\sigma^2(x)$. It is connected with the anomalous dimension of the mass operator, $\gamma_{\phi^2}$, listed in~\cite{Zerf:2017zqi,Gracey:2025aoj} as $\gamma_{\sigma^2} = 2\gamma_{\sigma} -2\gamma_{\phi^2}$.}
\begin{align}
  \label{eq:gamma-squared}
\gamma_{\sigma^2}(u, \lambda) &= 4Nu + \lambda - 2Nu^2 - 4Nu\lambda - \dfrac{5}{6}\lambda^2 + \left(178N^2 - \left(\dfrac{1131}{4} - 108\zeta_3\right)N\right)u^3 \\ \nonumber
&\quad - \left(12N^2 - \left(\dfrac{31}{2} + 60\zeta_3\right)N\right)u^2\lambda + \dfrac{11}{4}Nu\lambda^2 + \dfrac{7}{2}\lambda^3 + \ldots.
\end{align}

The set of leading-twist operators in this model can be classified according to the $SU(N)$ flavor symmetry. In particular, we consider:
\begin{itemize}
\item Flavor non-singlet operators
\begin{equation}
\label{eq:nonsinglet-operator-def}
\mathcal{O}^{q, A}_{s}(x) = \bar{q}_i(x)\slashed{n}\big(n\cdot \partial\big)^{s - 1}\tau^{A}_{ij}q_j(x), 
\end{equation}
  where $\tau^A$, ($A = 1, \ldots, N^2- 1$) is the generator of the $SU(N)$ group.
\item Flavor singlet operators~\footnote{Note the unusual normalization of the fermion singlet operators}
\begin{align}
\label{eq:singlet-operator-def}
\mathcal{O}_s^q(x) &= -s\sum_{i = 1}^N\bar{q}_i(x)\slashed{n}\big(n\cdot \partial\big)^{s - 1}q_i(x), \nonumber\\
\mathcal{O}_{s}^{\sigma}(x) &= \sigma(x)\big(n \cdot \partial\big)^s\sigma(x). 
\end{align}
\end{itemize}
Here $n$ is an auxiliary light-like vector ($n^2 = 0$). The operators~\eqref{eq:nonsinglet-operator-def} and~\eqref{eq:singlet-operator-def} require additional UV renormalization. The flavor non-singlet operator is multiplicatively renormalizable
\begin{equation}\label{eq:nsren}
  \big[\mathcal{O}^{q,A}_s(x)\big] = Z_{\text{ns}}(s)\mathcal{O}^{q,A}_s(x),
\end{equation}
while the flavor singlet operators mix
\begin{equation}\label{eq:sren}
\begin{pmatrix}[1.2]
    \big[\mathcal{O}^{q}_s(x)\big] \\
    \big[\mathcal{O}^{\sigma}_s(x)\big]
\end{pmatrix} = \begin{pmatrix}[1.2]
    Z_{qq}(s) & Z_{q\sigma}(s) \\ 
    Z_{\sigma q}(s) & Z_{\sigma\sigma}(s) 
\end{pmatrix}
\begin{pmatrix}[1.2]
\mathcal{O}_s^q(x) \\
\mathcal{O}_{s}^{\sigma}(x)
\end{pmatrix}.
\end{equation}
Here we omit the mixing with the total-derivative operators due to the block-diagonal structure of mixing. The corresponding RG equations take the form
\begin{align}
  \Big(\mu\partial_{\mu} + \beta_u\partial_u + \beta_\lambda\partial_\lambda + \gamma_{\text{ns}}(s)\Big)\big[\mathcal{O}^{q,A}_s(x)\big] &= 0,  \\ \left(\left(\mu\partial_{\mu} + \beta_u\partial_u + \beta_{\lambda}\partial_{\lambda}\right)\mathds{1}_2 + \begin{pmatrix}[1.2]
\gamma_{qq}(s) & \gamma_{q\sigma}(s) \\
\gamma_{\sigma q}(s) & \gamma_{\sigma\sigma}(s) 
\end{pmatrix}\right)\begin{pmatrix}[1.2]
    \big[\mathcal{O}^{q}_s(x)\big] \\
    \big[\mathcal{O}^{\sigma}_s(x)\big] 
\end{pmatrix} &= 0,
\end{align}
where $\mathds{1}_2$ is the identity $2\times 2$ matrix. Anomalous dimensions taking part in these equations are defined as
\begin{align}
  \label{eq:anomalous-dimensions-def}
\gamma_{\text{ns}}(s) = -\dfrac{d\ln Z_{\text{ns}}(s)}{d\ln \mu}, && \gamma_{ab}(s) = -\dfrac{dZ_{ac}(s)}{d\ln \mu}Z_{cb}^{-1}(s),
\end{align}
where $a, b \in \{q, \sigma\}$, and prior to this work were known to two-loop order~\cite{Manashov:2025kgf}.

\subsection{Critical behavior}

In $d = 4 - 2\epsilon$ dimensions the theory~\eqref{eq:gny-action-def} possesses several fixed points~(see~\cite{Zerf:2017zqi} for a detailed discussion), $u^*(\epsilon)$ and $\lambda^*(\epsilon)$, satisfying
\begin{align}
\label{eq:betas-zero}
\beta_u(u^*, \lambda^*) = 0, && \beta_{\lambda}(u^*, \lambda^*) = 0. 
\end{align}
Critical values of the couplings can then be found as series in $\epsilon$, and the three-loop $\beta$-functions~\eqref{eq:rg-functions} relevant for this work give an approximation up to $O(\epsilon^3)$. Solving~\eqref{eq:betas-zero} at one-loop order, we identify four critical points. Two of them are
\begin{align}
  \big(u^*, \lambda^*\big)_{0} = \big(0, 0\big), && \big(u^*, \lambda^*\big)_{\text{WF}} = \left(0, \,\, \dfrac{2\epsilon}{3}\right),
\end{align}
where the first critical point is trivial and the second corresponds to the Wilson-Fisher fixed point of the $\varphi^4$ theory, which is not relevant for the present discussion. Two other non-trivial critical points have the form
\begin{align}
\label{eq:nontrivial-fp}
  \big(u^*, \lambda^*\big)_{+} = \left(\dfrac{\epsilon}{3 + 2N},\,\,  \dfrac{3 - 2N + \delta}{9 + 6N}\epsilon\right), && \big(u^*, \lambda^*\big)_{-} = \left(\dfrac{\epsilon}{3 + 2N},\,\,  \dfrac{3 - 2N - \delta}{9 + 6N}\epsilon\right), 
\end{align}
where we use the convenient notation
\begin{align}
\delta = \sqrt{4N^2 + 132N + 9}, && \forall N \ge 0: \quad \delta \in \mathbb{R}.
\end{align}
Of the two non-trivial critical points~\eqref{eq:nontrivial-fp}, $\big(u^*, \lambda^*\big)_{+}$ is IR-stable, while $\big(u^*, \lambda^*\big)_{-}$ is the saddle point. This can be identified by studying the eigenvalues of the matrix of derivatives of the $\beta$ functions,
\begin{align}
\widehat{\omega} = \begin{pmatrix}
\partial_{u}\beta_u && \partial_u\beta_\lambda \\
\partial_\lambda\beta_u && \partial_\lambda\beta_\lambda 
\end{pmatrix}\Bigg|_{u = u^*, \lambda = \lambda^*}, && \omega_{1,2} \, -\, \widehat{\omega}\text{ eigenvalues}.
\end{align}
For the points~\eqref{eq:nontrivial-fp} we obtain 
\begin{align}
\omega^{+}_{1,2} = \left\{2\epsilon, \,\,\dfrac{2\delta}{3 + 2N}\epsilon\right\}, && \omega^{-}_{1,2} = \left\{2\epsilon, \,\,-\dfrac{2\delta}{3 + 2N}\epsilon\right\}.
\end{align}
In what follows, we focus entirely on the IR-stable critical point when referring to the critical behavior of the model~\eqref{eq:gny-action-def} and drop the second point, $(u^*, \lambda^*)_{-}$, from consideration.

Solving~\eqref{eq:betas-zero} further in perturbation theory gives the following expansion of the critical couplings
\begin{align}
  \label{eq:couplings-critical}
  u^*(\epsilon) = u_1\epsilon + u_2\epsilon^2 + u_3\epsilon^3 + O(\epsilon^4), && \lambda^*(\epsilon) = \lambda_1\epsilon + \lambda_2\epsilon^2 + \lambda_3\epsilon^3 + O(\epsilon^4),
\end{align}
where 
\allowdisplaybreaks{
\begin{subequations}
    \label{eq:u-critical}
    \begin{align}
      u_1 &= \dfrac{1}{3 + 2N},\\
      u_2 &= \dfrac{u_1^3}{108}\bigg[-8N^2 + 1032N + 441 + (4N + 66)\delta\bigg], \\
      u_3 &= \dfrac{u_1^5}{3888\,\delta}\bigg[-4416N^5 - 265120N^4 - 3506304N^3 - 4058208N^2 - 10140012N - 404190 \nonumber \\
      &\quad + \Big(2208N^4 - 12736N^3 + 656604N^2 + 1021680N + 412749 \nonumber \\
      &\quad - \big(419904N^2 + 1073088N + 664848\big)\zeta_3\Big)\delta\bigg].
\end{align}
\end{subequations}
\begin{subequations}
   \label{eq:lambda-critical}
\begin{align}
  \lambda_1 &= \dfrac{u_1}{3}\bigg[-2N + 3 + \delta\bigg], \\
  \lambda_2 &= \dfrac{u_1^3}{27\,\delta}\bigg[96N^4 + 1240N^3 + 5490N^2 - 1269N + 918 + \big(-48N^3 + 172N^2 + 1545N + 306\big)\delta\bigg], \\
  \lambda_3 &= \dfrac{u_1^5}{486\,\delta^3}\bigg[100608N^8 + 5916352N^7 + 82686688N^6 + 37493136N^5 - 205548408N^4 \nonumber \\
  &\quad - 87882516N^3  + 21816540N^2 + 24925239N + 516861 - \big(73728N^8 + 5985792N^7 \nonumber \\
  &\quad + 152326656N^6 + 1245995136N^5 + 2969374464N^4 + 2309402016N^3 + 285744672N^2 \nonumber \\
  &\quad  + 37318968N + 1889568\big)\zeta_3+ \Big(-50304N^7 - 2128160N^6 - 15865760N^5 - 9216696N^4 \nonumber \\
  &\quad + 61518456N^3 + 9467712N^2 + 2878740N + 172287  + \big(36864N^7 + 2384640N^6 \nonumber \\
  &\quad  + 41793408N^5 + 108832896N^4 + 67417920N^3 - 21520080N^2 - 10969992N \nonumber \\
  &\quad  - 629856\big)\zeta_3\Big)\delta\bigg].
\end{align}
\end{subequations}}

\begin{figure}[t]
  \centering
  \includegraphics[width=0.75\columnwidth]{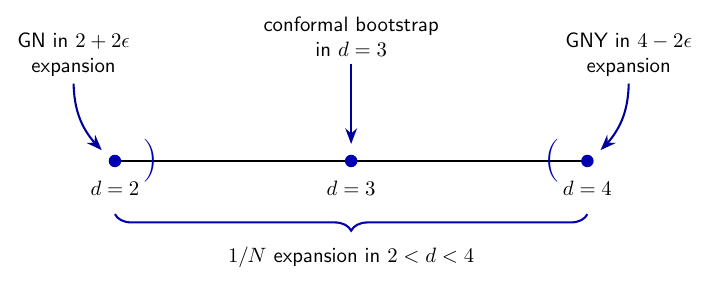}
  \caption{
    Different regimes where $\text{GN}$-universality class is studied. 
   }
    \label{fig:GN-universality}
\end{figure}

Considering the GNY model in $d = 4 - 2\epsilon$ dimensions at the critical values of the couplings, $(u^*, \lambda^*)_{+}$, gives an asymptotic $\epsilon$ expansion for the $d$-dimensional CFT, with $\epsilon$ playing the role of the perturbative expansion parameter. In particular, the results~\eqref{eq:u-critical} and~\eqref{eq:lambda-critical} allow us to extract the expansion up to $O(\epsilon^3)$. This fact can serve as a valuable cross-check for the leading-twist anomalous dimensions. On the one hand, the emergent conformal symmetry puts constraints on the analytical structure of anomalous dimensions~\eqref{eq:anomalous-dimensions-def} as functions of spin, see Section~\ref{subsec:reciprocity} and Section~\ref{sec:small-spin} for details.  
On the other hand, the critical exponents of the Gross-Neveu universality class are studied in different regimes (see Fig.~\ref{fig:GN-universality}), including the $\epsilon$ expansion in the $d = 2 + 2\epsilon$ Gross-Neveu model, the conformal bootstrap in three dimensions, and the $1/N$ expansion framework. In this paper, we focus on the latter and ensure that our results match the calculations of leading-twist operators in the $1/N$ expansion~\cite{Giombi:2017rhm, Manashov:2016uam}, see Section~\ref{subsec:large-N}.

\subsection{Generalized Lagrangian and emergent SUSY\label{subsec:genl-and-susy}}

It is well known that the symmetry of the GNY model~\eqref{eq:gny-action-def} continued to the number of flavors $N = 1/4$ is enhanced by $\mathcal{N} = 1$ supersymmetry, see~\cite{Grover:2012bm,Fei:2016sgs} for details. This fact manifests itself, for example, in the coincidence of the scheme-independent critical anomalous dimensions of the fermion, $\gamma_{q}$, and the scalar field, $\gamma_{\sigma}$:~\footnote{Note that this relation holds when $d = 3$. This does not make any difference for the anomalous dimensions~\eqref{eq:rg-functions} at order $O(\epsilon^3)$, but in principle has to be taken into account. We dicuss this in more details in Section~\ref{sec:res}.}
\begin{align}
  \label{eq:susy-anomalous-dimensions}
  \text{at } N = 1/4: \quad \gamma^*_{q}(\epsilon) = \gamma^{*}_{\sigma}(\epsilon), && \gamma_{q,\sigma}^*(\epsilon) = \gamma_{q, \sigma}(u^*(\epsilon), \lambda^*(\epsilon)).
\end{align}   
One can explicitly check this property using~\eqref{eq:rg-functions} and~\eqref{eq:couplings-critical}. In what follows we describe an approach introduced in~\cite{Chakraborty:2026lna}, which transforms this emergent supersymmetry into a practical tool for calculations in the (non-supersymmetric) GNY model.

%
We define the following generalized Lagrangian
\begin{align}
\label{lgen}
 \mathcal{L}^{\text{gen}} &= \frac{1}{2}\partial_\mu \phi^I\partial^\mu \phi_I+ \bar\psi^i\slashed{\partial}\psi_i+g(\bar Z^I)^{ij}\phi_I\psi_i\psi_j + g( Z^I)_{ij}\phi_I\bar\psi^i\bar\psi^j  + \frac{\xi}{4!}M^{IJKL}\phi_I\phi_J\phi_K\phi_L
\, ,
\end{align}
where $\phi_I$ are real scalars and $\psi_i$ are Weyl fermions transforming in the fundamental representations of their respective global $SO(N_s)$ and $SU(2N_f)$ flavor groups. Note that in the fermionic sector we use $2N_f$ as the flavor number. This is needed to arrange the Weyl fermions into Dirac fermions with $SU(N_f)$ symmetry --- the usual flavor content seen in the GNY model. The \emph{flavor elements} $(\bar Z^I)^{ij}, (Z^I)_{ij}$, and $M^{IJKL}$ are abstract algebraic structures, whose definition is based on the following constraints on the theory:
\begin{itemize}
    \item \textbf{Classical condition:} The Lagrangian is real-valued and transforms as a singlet under all symmetries.
    \item \textbf{Quantum condition:} No evanescent flavor elements emerge in the renormalized Lagrangian from loop corrections.
\end{itemize}
It was proved in \cite{Chakraborty:2026lna} that the quantum condition holds if and only if 
\begin{equation}\label{eq:flavclifford}
    Z^I\bar Z^J+Z^J\bar Z^I = 2\delta^{IJ}\;,
\end{equation}
which then implies 
\begin{equation}\label{mijkl}
    M^{IJKL} = \frac{1}{3}(\delta^{IJ}\delta^{KL}+\delta^{IK}\delta^{JL}+\delta^{IL}\delta^{JK}).
\end{equation}
The above relations by themselves are sufficient to evaluate traces and index contractions of the flavour elements, and allow us to compute the flavour factors of all relevant one-particle irreducible (1PI) correlators as polynomials in the flavour dimensions $(N_s, N_f)$.

The generalized Lagrangian corresponds to different theories depending on the choice of $(N_s, N_f)$. In particular,~\eqref{lgen} contains the Gross-Neveu-Yukawa, Nambu–Jona-Lasinio–Yukawa, and Wess-Zumino models, see Table~\ref{tab:models-text}. Indeed, one can easily see that the matrix representations for $Z^I,\bar Z^I$ corresponding to each value of $(N_s,N_f)$ give us the classical Lagrangians for the mentioned models.
\begin{table}[ht]
    \centering
    {\renewcommand{\arraystretch}{1.3}
    \begin{tabular}{c|c}
    \hline
    $(N_s,N_f)$ & Model \\
    \hline
    \renewcommand{\arraystretch}{0.5}
    $(1,N)$ & Gross-Neveu-Yukawa \\
    $(2,N)$ & Nambu–Jona-Lasinio–Yukawa \\
    $(2,1/2)$ & 4D $\mathcal{N}=1$ Wess-Zumino \\
    $(1,1/4)_{d=3}$ & 3D $\mathcal{N}=1$ Wess-Zumino \\
    \hline
    \end{tabular}}
    \caption{Models corresponding to specific representations of Eq.~\eqref{eq:flavclifford} for the generalized Lagrangian, $\lgen$. The subscript $d=3$ denotes that we consider the Lagrangian in three spacetime dimensions instead of four.}
    \label{tab:models-text}
\end{table}

For generic values of $N_s$ and $N_f$, Lagrangian~\eqref{lgen} describes a non-supersymmetric theory, but for specific values supersymmetry emerges and connects anomalous dimensions of the scalar and fermionic fields, as well as the $\beta$ functions for the respective Yukawa and scalar couplings, $u$ and $\lambda$. Relations~\eqref{eq:susy-anomalous-dimensions} are exactly the consequence of this phenomenon at $N_s = 1, N_f = 1/4$. In~\cite{Chakraborty:2026lna} it was shown that the only other supersymmetric point is $N_s = 2, N_f = 1/2$. At the points where supersymmetry emerges, we can use additional Ward identities to connect Green's functions involving different operators and, therefore, corresponding Feynman integrals. Remarkably, this connection can then be continued to the case of generic $N_s$ and $N_f$. We describe the details of this approach in Section~\ref{sec:opmitsusy}.

\section{Calculations and Results\label{sec:results}}
\subsection{General methodology\label{subsec:method}}

The anomalous dimensions of the operators are obtained by computing and renormalizing the 
forward off-shell amputated (1PI) Green's functions for the operator matrix elements (OMEs). 
For example, for the non-singlet operator we have
\begin{equation}
  \label{eq:green-quark-ns}
    \Gamma_{ij}^{q, A}(Q,s) \coloneq \int d^dx\;\langle q_i|\mathcal{O}^{q, A}_{s}(x)|q_j
    \rangle_{\text{1PI}},  
\end{equation}
where $|q_i\rangle$ is a fermion state with (off-shell) momentum $Q^\mu$ and flavor index 
$i$. Note that~\eqref{eq:green-quark-ns} defines the bare (unrenormalized) Green's function 
since we insert the bare operator. Calculating the divergent part of the Green's 
function~\eqref{eq:green-quark-ns}, we fix the local $\overline{\text{MS}}$ counterterm 
$Z_{\text{ns}}(s)$, see~\eqref{eq:nsren}. Due to the mixing present in the singlet sector, 
we consider the following four forward off-shell OMEs:
\begin{align}
  \label{eq:green-singlet}
        \Gamma^{qq}(Q,s)& \coloneq  \frac{1}{N}\sum_{i = 1}^{N}\int d^dx\;\langle q_i|
        \mathcal{O}^{q}_{s}(x)|q_i\rangle_{\text{1PI}} \nonumber\\
        \Gamma^{q\sigma}(Q,s) &\coloneq  \int d^dx\;\langle \sigma|\mathcal{O}^{q}_{s}(x)|
        \sigma\rangle_{\text{1PI}} \nonumber\\
        \Gamma^{\sigma q}(Q,s) &\coloneq  \frac{1}{N}\sum_{i = 1}^{N}\int d^dx\;\langle q_i|
        \mathcal{O}^{\sigma}_{s}(x)|q_i\rangle_{\text{1PI}}\nonumber\\
        \Gamma^{\sigma\sigma}(Q,s) &\coloneq  \int d^dx\;\langle \sigma|\mathcal{O}^{\sigma}_{s}
        (x)|\sigma\rangle_{\text{1PI}}
\end{align}
where $|\sigma\rangle$ is a scalar state with momentum $Q^\mu$. The divergent part of each 
individual bare $\Gamma^{ab}$ defines the local counterterm $Z_{ab}(s)$, see 
Eq.~\eqref{eq:sren}, in the $\overline{\text{MS}}$ scheme. The corresponding anomalous 
dimensions in both the singlet and non-singlet cases are then calculated using the 
definition~\eqref{eq:anomalous-dimensions-def}.

In practice, we perform calculations in the generalized model defined by~\eqref{lgen} with 
arbitrary flavor content, $(N_s, N_f)$. This generalization does not create any notable 
difference in complexity, since it requires only knowledge of additional flavor-symmetry 
factors. As a result, however, we can apply the optimization based on emergent SUSY, see 
Section~\ref{sec:opmitsusy} for details, which saves days of calculations.

On the technical side, we generate the diagrams for the Green's 
functions~\eqref{eq:green-quark-ns} and~\eqref{eq:green-singlet} with 
\texttt{Qgraf}~\cite{Nogueira:1991ex}. The operator insertions are defined as external 
bosonic source fields in the \texttt{Qgraf} file. The resulting diagram files are 
subsequently handled in \texttt{FORM}~\cite{Vermaseren:2000nd,Kuipers:2012rf,Ruijl:2017dtg,
Davies:2026cci}. Within this framework, we implement the relevant Feynman rules, classify 
the diagram topologies, and evaluate the flavor and spinor structures. In addition, 
\texttt{FORM} translates the \texttt{Qgraf} output into the input format required by the 
database framework \texttt{Minos}~\cite{minos}, which is used to organize the subsequent 
diagrammatic calculation. The management of diagram topologies, flavor structures, and 
individual diagram contributions is performed within \texttt{Minos}. For the evaluation of 
loop integrals, \texttt{Minos} is interfaced with \texttt{FORCER}~\cite{Ruijl:2017cxj}, 
allowing the required massless two-point integrals to be computed through four loops. The 
association of the resulting \texttt{FORCER} integrals with the appropriate flavor factors, 
as well as the reconstruction of the complete propagator and vertex expressions, is carried 
out automatically by \texttt{Minos}.

In the above-mentioned procedure we are unable to treat the spin variable, $s$, analytically. The typical output of such calculations reads
\begin{align}
  \label{eq:output-example}
  \gamma_{\text{ns}}(2) &= -\frac{29 }{216}\lambda^2 u + \frac{14 }{9}\lambda u^2 + 
  u^3\left(-\frac{2533}{1944} + \frac{200 N}{27} - \frac{82 N^2}{27} + \frac{10 \zeta_3}
  {3}\right),\nonumber\\
  \gamma_{\text{ns}}(4) &= -\frac{79 }{400}\lambda^2 u + \frac{607 }{300}\lambda u^2 + 
  u^3\left(-\frac{2371273}{2400000} + \frac{1332151 N}{120000} - \frac{3321 N^2}{1000} + 
  \frac{303 \zeta_3}{100}\right), \nonumber\\ 
  \gamma_{\text{ns}}(6) &= -\frac{565}{2646} \lambda^2 u + \frac{4559}{2205} \lambda u^2 
  + u^3\left(-\frac{1594579421}{1633640400} + \frac{152449639 N}{12965400} - \frac{29800 
  N^2}{9261} + \frac{442 \zeta_3}{147}\right),\nonumber \\
  &\ldots.
\end{align}
It is well known that the leading-twist anomalous dimensions as functions of $s$ consist of 
harmonic sums~\cite{Vermaseren:1998uu}
\begin{equation}
  \label{eq:harmsums}
  S_{m_1,\dots,m_k}(s)=\sum_{i=1}^{s}
  \frac{\operatorname{sgn}(m_1)^{\,i}}{i^{|m_1|}}\,S_{m_2,\dots,m_k}(i),
  \qquad S_\varnothing=1.
\end{equation}
Moreover, the transcendental weight of these sums, $w = |m_1| + \ldots + |m_k|$, at 
$\ell$-th loop level should not exceed $2\ell - 1$.~\footnote{Note that this bound is usually discussed in the QCD context. For the GNY model we observe 
a drop in the highest transcendentality by roughly one loop order, see Section~\ref{sec:res} 
and the two-loop result~\cite{Manashov:2025kgf}.} Each harmonic sum of transcendental 
weight $w$ can be accompanied by a factor $1/(s + a)^k$, where $w + k \le 2\ell + 1$ and 
$a$ is an integer. Based on the discussion in Section~\ref{subsec:reciprocity} and 
Section~\ref{sec:small-spin}, we expect only $a = 0$ and $a = 1$ to appear. These 
constraints allow us to effectively transform the output~\eqref{eq:output-example} into an 
analytical expression in $s$ by constructing a suitable ansatz and determining the 
corresponding coefficients.

A remarkable feature of this procedure is that the size of the ansatz is significantly 
larger than the number of available equations. For example, the number of independent 
harmonic sums at a given transcendental weight $w$ is $2\cdot 3^{w - 1}$, which at three 
loops gives $162$, not taking into account rational functions. As \texttt{FORCER} can only 
deal with scalar products of loop and external momenta, the light-cone projected Feynman 
rules need to be converted into a \texttt{FORCER}-compatible form using harmonic 
projections \cite{Larin:1996wd}, which leads to a rapid increase of terms at higher values 
of spin. As a result, usually only a relatively modest amount of data points is feasible, 
$s \lessapprox 20$. Fortunately, the set of equations for each coefficient in~\eqref{eq:output-example} 
(in $\lambda$, $u$, $N$, $\zeta_3$) only includes rational numbers as coefficients, which transforms 
the problem of analytical reconstruction into a problem of solving a set of Diophantine 
equations. Such equations can be effectively solved using the LLL 
algorithm~\cite{Lenstra:1982eee}. In particular, we use the built-in \texttt{Mathematica} 
method \texttt{LatticeReduce}.

\subsection{Optimisation with emergent SUSY}\label{sec:opmitsusy}
In this section we describe the procedure that allows us to turn the emergent SUSY 
outlined in Section~\ref{subsec:genl-and-susy} into a practical tool for calculations. 
Such an optimization is particularly useful in view of the growth in complexity of 
calculations for large values of spin. Here we provide a short explanation suitable for 
practical implementation, for additional details see~\cite{Chakraborty:2026lna}.

In a theory which possesses supersymmetry one can derive corresponding SUSY Ward identities. 
In particular, in the singlet sector of the generalized theory~\eqref{lgen}, we expect the 
following relation to hold
\begin{equation}
  \label{bare_ward}
    \frac{2}{s}\Gamma^{q\sigma}(s) + 2\,\Gamma^{\sigma\sigma}(s) = 2\,\Gamma^{qq}(s) + 
    2s\,\Gamma^{\sigma q}(s),
    \end{equation}
when $(N_s, N_f)$ corresponds to one of the two possible SUSY points, see 
Table~\ref{tab:models-text}. Note that here we use the normalization specific to the 
definitions~\eqref{eq:green-singlet}. This relation can be translated into a relation among 
Feynman integrals. The method can be summarized as follows:
\begin{enumerate}
\item Express the bare Green's functions $\Gamma^{ab}$ of the singlet operators as a sum 
of flavor factors $f_i(N_s,N_f)$ multiplied by flavor-independent loop integrals $I_i$:
\begin{equation}\label{eq:gencor}
    \Gamma^{ab} = \sum_{k + m = \ell}\sum_{i} u^m \lambda^{k} f_i(N_s,N_f)\;I^{ab}_i(s),
\end{equation}
where $\ell$ corresponds to the desired loop order and each integral implicitly depends on $m$ and $k$.

\item Substitute the emergent SUSY points: $u = 3\lambda$, $(N_s,N_f) = (2,\,1/2)$ and 
$(1,\,1/4)$. Under this substitution, Eq.~\eqref{bare_ward} holds and yields a set of 
relations among the loop integrals.

\item Employ these relations to eliminate the most technically demanding integrals from 
the general expressions for the $\Gamma^{ab}$'s.

\item The remaining calculated integrals can then be substituted back into 
expression~\eqref{eq:gencor}, yielding results for generic $N_s$ and $N_f$.
\end{enumerate}

At higher moments, the computational time is almost entirely dominated by diagrams of the 
following non-planar topology:
\begin{center}
\includegraphics[width=0.3\linewidth]{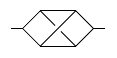}    
\end{center}
This is because the complexity arising from numerators increases much faster in non-planar 
diagrams than in planar ones. Remarkably, the diagrams of this topology satisfy the SUSY 
Ward identity \emph{independently} of the other topologies, and hence we obtain the 
following relation
\begin{equation}
        N_s(N_s^2-6N_s+4)\Big(\,I_{\text{non-planar}}^{qq} + s\,I_{\text{non-planar}}^{
        \sigma q}\Big) \underset{\scriptstyle\text{SUSY}}{=} N_f(N_s^2-6N_s+4)\Big(\frac{1}
        {s}I_{\text{non-planar}}^{q\sigma} + \,I_{\text{non-planar}}^{\sigma\sigma}\Big)\;,
\end{equation}
where we have written the non-planar contributions from each $\Gamma^{ab}$ in the form of 
Eq.~\eqref{eq:gencor}. Note that each $I_{\text{non-planar}}^{qq}$ is a sum of integrals 
in the non-planar topology having the same flavor factor. Substituting the two emergent 
SUSY points gives an equation between the integrals similar to~\eqref{bare_ward}:
\begin{equation}
    4\Big(I_{\text{non-planar}}^{qq} + s\,I_{\text{non-planar}}^{\sigma q}\Big) = \frac{1}{s}
    I_{\text{non-planar}}^{q\sigma} + \,I_{\text{non-planar}}^{\sigma\sigma},
\end{equation}
which can be used to eliminate one of the four integral families. We choose to eliminate 
$I_{\text{non-planar}}^{qq}$ as it takes the longest time of the four, yielding an 
approximately 25\% speed-up, as confirmed by explicit calculation. The total calculation 
of all even moments up to $s = 30$, which was projected to take a little more than 14 days 
with \texttt{TFORM} running on 32 cores, took a little less than 11 days with this 
technique.

\subsection{Results}\label{sec:res}

In this section we present the results of our calculations. As discussed in 
Section~\ref{subsec:method}, calculations are done for the theory described by the 
generalized Lagrangian~\eqref{lgen}. For brevity, we do not reproduce here the results for 
generic values of $(N_s, N_f)$, which can be found in Appendix~\ref{app:lgen}. To obtain 
results for the GNY model~\eqref{eq:gny-action-def}, one needs to substitute $N_s = 1$, 
$N_f = N$. Note that using the results of Appendix~\ref{app:lgen} one can also easily 
obtain the leading-twist anomalous dimensions in the singlet sector of NJLY model as well as the Wess-Zumino 
model, see Table~\ref{tab:models-text}.

In what follows we provide the results for the non-singlet anomalous dimension and singlet matrix of anomalous dimensions. In both cases the results are written in terms of perturbative expansion
\begin{equation}
  \gamma(s) = \gamma_1(s) + \gamma_2(s) + \ldots,
\end{equation}  
where the index $\ell$ in $\gamma_{\ell}$ represents the contribution at $\ell$-th loop order. The notation $\gamma_{q, \ell}$ and $\gamma_{\sigma, \ell}$ follows the similar logic. Here we omit the variable in the harmonic sums, $S_{m_1, \ldots, m_k} \equiv S_{m_1, \ldots, m_k}(s)$. In this section we also provide the results 
for one- and two-loop orders for completeness. The original references for these 
are~\cite{Giombi:2017rhm, Manashov:2025kgf}. For the reader's convenience, we provide all expressions in the ancillary file. 

\paragraph{Non-singlet operator:}

Here we present results for the non-singlet operator~\eqref{eq:nonsinglet-operator-def}
\begin{equation}
  \label{eq:ns-results}
  \gamma_{\text{ns}}(s) = \gamma_{\text{ns},1}(s) + \gamma_{\text{ns}, 2}(s) + \gamma_{\text{ns}, 3}(s)+\ldots.
\end{equation}
The anomalous dimension, $\gamma_{\text{ns}}(s)$, is naturally defined for $s \in \mathbb{Z}_{> 0}$. The results for the perturbative contributions read 
\begin{subequations}
  \label{eq:gny-ns}
\begin{align}
  \gamma_{\text{ns}, 1}(s) &= 2\gamma_{q, 1} - \dfrac{2u}{s(s + 1)},\\
  \gamma_{\text{ns}, 2}(s) &= 2\gamma_{q, 2} -\frac{2u^2}{s(1+s)}\bigg[3S_1 
  +\frac{1+2s}{s^2(1+s)^2} + \frac{1+5s+2N(1+2s)}{s(1+s)} - 8 - 6N \nonumber \\&\quad 
  + \left(1-(-1)^s\right)\left(1-\frac{1}{s(1+s)}\right)\bigg],\\
\gamma_{\mathrm{ns}, 3}(s) &= 
2\gamma_{q, 3} + u\lambda^2\biggl[ -\frac{2 s+1}{6 s^2 (s+1)^2} +\frac{17}{24 s (s+1)} 
\biggr] +2\,u^2\lambda\biggl[ \frac{4}{s (s+1)} S_1 +\frac{2 (2 s+1)}{s^2 (s+1)^2} 
-\frac{9}{s (s+1)} \biggr] \notag\\
&\quad - \dfrac{u^3}{s(s + 1)}\Bigg[\frac{8}{s(1+s)}S_{3} + \frac{16}{s(1+s)}
\left(S_{1,-2} - \dfrac{1}{2}S_{-3}\right) -\bigg(\frac{4}{s^2(1+s)^2} 
+\frac{8}{s(1+s)}- 8\bigg)S_{-2} \nonumber\\ &\quad +9\,S_1^2+\bigg(\frac{6}{s(1+s)}
+12N+15\bigg)S_{2}+\bigg(\frac{12+24s}{s^2(1+s)^2}+\frac{12N+24Ns+30s}{s(1+s)}
-50N-\frac{69}{2}\bigg)S_{1}  \nonumber \\ &\quad -\Big(\frac{24}{s(1+s)} - 12\Big)
\zeta_3+\frac{4}{s^4(1+s)^4}+\frac{12N+16-30s}{s^3(1+s)^3}+\frac{8N^2+20N-88Ns+30
-106s}{s^2(1+s)^2} \nonumber \\ &\quad -\frac{96N^2s - 68N + 316Ns - 156 + 157s}
{2s(1+s)}+\dfrac{128N^2 + 976N + 99}{8} \nonumber \\ &\quad
-(-1)^s\bigg\{\Big(\frac{4}{s(1+s)}-4\Big)S_{3} + \Big(\frac{8}{s(1+s)}-8\Big)
\left(S_{1,-2} - \dfrac{1}{2}S_{-3}\right)+\Big(\frac{4}{s^2(1+s)^2} \nonumber \\ 
&\quad + \frac{12}{s(1+s)}+8\Big)S_{-2}+\Big(\frac{-8N-2}{s(1+s)}+8N+6\Big)S_{1}
+\Big(-\frac{12}{s(1+s)}+12\Big)\zeta_3 -\frac{4 + 12s}{s^3(1+s)^3} \nonumber \\ 
&\quad +\frac{-4N-16Ns+8-12s}{s^2(1+s)^2}+\frac{24N+8Ns+46+10s}{s(1+s)}-16N-22
\bigg\}\Bigg].
\end{align}
\end{subequations}

\paragraph{Singlet operators:}

Here we present the anomalous dimension matrix for the singlet 
operators~\eqref{eq:singlet-operator-def}
\begin{equation}
  \label{eq:singlet-results}
  \widehat{\gamma}(s) = \begin{pmatrix}
    \gamma_{qq,1}(s) & \gamma_{q\sigma, 1}(s) \\
    \gamma_{\sigma q,1}(s) & \gamma_{\sigma \sigma, 1}(s)
  \end{pmatrix} + 
  \begin{pmatrix}
    \gamma_{qq,2}(s) & \gamma_{q\sigma, 2}(s) \\
    \gamma_{\sigma q,2}(s) & \gamma_{\sigma \sigma, 2}(s)
  \end{pmatrix} + \begin{pmatrix}
    \gamma_{qq,3}(s) & \gamma_{q\sigma, 3}(s) \\
    \gamma_{\sigma q,3}(s) & \gamma_{\sigma \sigma, 3}(s)
  \end{pmatrix} + \ldots 
\end{equation}
The singlet matrix as function of spin, $\widehat{\gamma}(s)$ is naturally defined 
only for even values of spin, $s \in 2\mathbb{Z}_{>0}$. The results for entries 

\begin{subequations}
\label{eq:gny-full-qq}
\begin{align}
  \gamma_{qq,1}(s) &= 2\gamma_{q, 1} - \dfrac{2u}{s(s + 1)}, \\
  \gamma_{qq,2}(s) &= 2\gamma_{q, 2} - \dfrac{2u^2}{s(s + 1)}\bigg[3\,S_{1}+\frac{1+2s}
  {s^2(1+s)^2}+\frac{4Ns+6N+5s+1}{s(1+s)}-10N-8\bigg],\\
\gamma_{qq, 3}(s) &= 
2\gamma_{q, 3} - \dfrac{u\lambda^2}{s(s + 1)}\biggl[\frac{2 s+1}{6 s(s+1)} -\frac{17}{24} 
\biggr] + \dfrac{2\,u^2\lambda}{s(s + 1)}\biggl[4S_1 +\frac{2 (2 s+1)}{s(s+1)} -9 
\biggr] \notag\\
&\quad - \dfrac{u^3}{s(s + 1)}\Bigg[\bigg(\frac{4}{s(1+s)}+4\bigg)S_{3}+\bigg(\frac{8}
{s(1+s)}+8\bigg)\bigg(S_{1,-2}-\frac12 S_{-3}\bigg) - \bigg(\frac{8}{s^2(1+s)^2} 
\nonumber \\ &\quad  + \frac{20}{s(1+s)}\bigg)S_{-2}+\big(8N^2-2N+9\big)S_1^2+\bigg(
\frac{6}{s(1+s)}+8N^2+34N+15\bigg)S_{2} \nonumber \\ &\quad  +\bigg(\frac{12+24s}
{s^2(1+s)^2}+\frac{16N^2s+32N^2+32Ns+60N+30s+2}{s(1+s)}-64N^2-90N-\frac{81}{2}
\bigg)S_{1} \nonumber \\ &\quad  -\frac{12}{s(1+s)}\zeta_3+\frac{4}{s^4(1+s)^4}+
\frac{48Ns+44N-18s+20}{s^3(1+s)^3}\nonumber \\ &\quad + \frac{48N^2s+24N^2-4Ns+72N
-94s+22}{s^2(1+s)^2} \nonumber \\ &\quad -\frac{224N^2s+160N^2+492Ns+220N+177s-64}
{2s(1+s)}+\frac{1152N^2+1680N+275}{8}\Bigg].
\end{align}
\end{subequations}

\begin{subequations}
    \label{eq:gny-full-qs}
\begin{align}
  \gamma_{q\sigma, 1}(s) &= -8Nu,\\
  \gamma_{q\sigma,2}(s) &= -8Nu^2\bigg[3\,S_{1}+\frac{1+2s}{s^2(1+s)^2}+\frac{3}
  {2s(1+s)}-7\bigg],\\
  \gamma_{q\sigma,3}(s) &= -\dfrac{8Nu}{s(s + 1)}\lambda^2\bigg[\dfrac{3 + 2s}{4s(s + 1)} 
  - 1\bigg] + 8Nu^2\lambda \bigg[2S_1 + \dfrac{2}{s(s + 1)} - 7\bigg] \nonumber \\&\quad 
  - 8Nu^3\bigg[\frac{2}{s(1+s)}S_{3}+\frac{4}{s(1+s)}\Big(S_{1,-2}-\frac12 S_{-3}\Big) 
  - \Big(\frac{4}{s^2(1+s)^2} + \frac{14}{s(1+s)}\Big)S_{-2} \nonumber \\ &\quad 
  +\bigg(\frac32N+\frac{33}{8}\bigg)S_1^2+\Big(\frac{3}{s(1+s)}+\frac32N+\frac{69}{8}
  \Big)S_{2}+\Big(\frac{12+21s}{2s^2(1+s)^2}+\frac{1}{s(1+s)} \nonumber \\ &\quad 
  -13N-\frac{87}{4}\Big)S_{1} - \Big(\frac{6}{s(1+s)} - 6\Big)\zeta_3 + \frac{2}
  {s^4(1+s)^4}+\frac{12Ns+10N+11}{s^3(1+s)^3}\nonumber \\ &\quad -\frac{152Ns+76N+114s
  +9}{8s^2(1+s)^2}+\frac{N-13}{s(1+s)}+\frac{940N+269}{32}\bigg].
\end{align}
\end{subequations}

\begin{subequations}
\label{eq:gny-full-sq}
\begin{align}
  \gamma_{\sigma q,1}(s) &= -\dfrac{2u}{s(s + 1)},\\ 
  \gamma_{\sigma q,2}(s) &= -\dfrac{2u^2}{s(s + 1)}\bigg[2(1+2N)\,S_{1}+\frac{1+2s}
  {s^2(1+s)^2}+\frac{8Ns+10s+7}{2s(1+s)}-8N-5\bigg],\\
  \gamma_{\sigma q,3}(s) &= -\dfrac{u\lambda^2}{3s(s + 1)}\bigg[S_1 + \dfrac{6s + 3}
  {2s^2(s + 1)^2} + \dfrac{1}{s + 1} - \dfrac{13}{4}\bigg] + \dfrac{2u^2\lambda}
  {s(s + 1)}\bigg[2S_1 + \dfrac{4s + 2}{s(s + 1)} - 7\bigg] \nonumber \\ 
  &\quad -\dfrac{u^3}{s(s + 1)}\bigg[\frac{4}{s(1+s)}S_{3}+\frac{8}{s(1+s)}
  \Big(S_{1,-2}-\frac12 S_{-3}\Big)- \bigg(\frac{8}{s^2(1+s)^2} + \frac{16N + 24}
  {s(1+s)}\bigg)S_{-2} \nonumber \\ &\quad +\bigg(\frac{-4N+1}{2s(1+s)}+12N^2+15N+
  \frac92\bigg)S_1^2+\bigg(\frac{12N+9}{2s(1+s)}+12N^2+27N+\frac{21}{2}\bigg)S_{2} 
  \nonumber \\ &\quad +\bigg(\frac{8N+12Ns+10+21s}{s^2(1+s)^2}+\frac{24N^2s+36N+54Ns
  +11+21s}{s(1+s)}-48N^2-96N-26\bigg)S_{1} \nonumber \\ &\quad -\Big(\frac{12}{s(1+s)} 
  - 12\Big)\zeta_3 + \frac{4}{s^4(1+s)^4}+\frac{20N+12Ns+28-3s}{s^3(1+s)^3} \nonumber 
  \\ &\quad +\frac{-96N^2s-20N-488Ns+193-144s}{4s^2(1+s)^2}  +\frac{48N^2-96N^2s-12N
  -292Ns-7-127s}{2s(1+s)}\nonumber \\ &\quad +\frac{1684N+239}{16}\bigg].
\end{align}
\end{subequations}

\begin{subequations}
 \label{eq:gny-full-ss}
\begin{align}
\gamma_{\sigma\sigma, 1}(s) &= 2\gamma_{\sigma,1}, \\
\gamma_{\sigma\sigma, 2}(s) &= 2\gamma_{\sigma, 2} -\dfrac{\lambda^2}{s(s + 1)} - 
\dfrac{8Nu^2}{s(s + 1)}\bigg[\dfrac{1 + 2s}{s(s + 1)} - 2\bigg], \\
\gamma_{\sigma\sigma, 3}(s) &= 2\gamma_{\sigma,3} - \dfrac{\lambda^3}{s(s + 1)}\bigg[
2S_1 + \dfrac{6s + 3}{2s(s + 1)} - 5\bigg] - \dfrac{4Nu\lambda^2}{s(s + 1)}\bigg[S_1 
+ \dfrac{1}{s + 1} - 3\bigg] \nonumber \\ &\quad  + \dfrac{8Nu^2\lambda}{s(s + 1)}\bigg[
6S_1 + \dfrac{6s + 3}{s(s + 1)}-8\bigg] - \dfrac{8Nu^3}{s(s + 1)}\bigg[2\,S_{3}+4
\Big(S_{1,-2}-\frac12 S_{-3}\Big) \nonumber \\ &\quad +\Big(4-\frac{4}{s(1+s)}\Big)
S_{-2} - \Big(N - \frac14\Big)S_1^2+\Big(3N+\frac94\Big)S_{2}+\Big(\frac{6Ns+4N+9s+5}
{s(1+s)} \nonumber \\ &\quad -4N-7\Big)S_{1}-6\,\zeta_3+\frac{2}{s^3(1+s)^3} - 
\frac{12Ns + 4N- 15s - 30}{2s^2(1+s)^2} - \frac{12Ns + 31s + 4}{s(1+s)}+30\bigg].
\end{align}
\end{subequations}

Note that in the singlet sector the eigenvalues of the anomalous dimension matrix play a distinguished physical role:
\begin{equation}
  \label{eq:eigenvalues-definition}
  \gamma_{\pm}(s) = \dfrac{1}{2}\Big(\operatorname{tr}\widehat{\gamma}(s) \pm 
  \sqrt{\big(\operatorname{tr}\widehat{\gamma}(s)\big)^2 - 
  4\operatorname{det}\widehat{\gamma}(s)}\Big),
\end{equation}
and we use $\gamma_{\pm}(s)$ in the analysis of critical behavior.

\paragraph{Three-dimensional enhancement:} 

In Section~\ref{subsec:genl-and-susy} it was mentioned that at $N = 1/4$ the GNY model 
possesses supersymmetry associated with the supersymmetry of the 3D 
$\mathcal{N} = 1$ Wess-Zumino model. This statement, however, should be treated with care, 
as we originally formulate our model~\eqref{eq:gny-action-def} in four dimensions. In 
particular, we treat all traces of an odd number of Dirac matrices as zero. This is no 
longer true in the 3D Clifford algebra. In~\cite{Zerf:2017zqi} it was 
observed that the supersymmetry at $N = 1/4$ does not hold at four loops, unless one 
considers additional contributions from traces of an odd number of Dirac matrices.

For the singlet operator Ward identity~\eqref{bare_ward}, this effect is already observed 
at three loops, as was pointed out in~\cite{Chakraborty:2026lna}, where there exists a 
surviving odd-trace contribution in $\gamma_{qq}$:
\begin{equation}
    \gamma_{qq, 3}(s)\big|_{d = 3} = \gamma_{qq, 3}(s)\big|_{d = 4} -16\,u^3N
    \left(\frac{1}{s(s+1)}-\frac{2}{s^2(s+1)^2}\right)\left(2S_{-2}+S_1\right). 
\end{equation}
In order to reproduce the SUSY-based relations, one has to take this contribution into 
account. The other entries of the anomalous dimension matrix, as well as the non-singlet 
anomalous dimensions, are not affected at this loop order.

\section{Consistency checks\label{sec:checks}}
\subsection{Conserved currents}
The model~\eqref{eq:gny-action-def} has two conserved currents relevant to the leading-twist operators. The $SU(N)$ symmetry implies the conservation of
\begin{equation}
  J^A_{\mu}(x) = \bar{q}_{i}(x)\gamma_{\mu}\tau^A_{ij}q_j(x).
\end{equation}
Another relevant conserved current is the stress-energy tensor
\begin{equation}
  T_{\mu\nu}(x) = \dfrac{1}{2}\bar{q}(x)\gamma_{\mu}\partial_{\nu}q(x) + (\mu \leftrightarrow \nu) + \partial_{\mu}\sigma\partial_{\nu}\sigma - \delta_{\mu\nu}\mathcal{L}.
\end{equation}
The light-cone projections of these currents can be represented as special combinations of operators~\eqref{eq:nonsinglet-operator-def} and~\eqref{eq:singlet-operator-def}
\begin{align}
  n^{\mu}J^{A}_{\mu}(x) = \mathcal{O}_{s = 1}^{q, A}(x), && n^{\mu}n^{\nu}T_{\mu\nu}(x) = -\dfrac{1}{2}\mathcal{O}^q_{s = 2}(x) - \mathcal{O}_{s = 2}^{\sigma}(x).
\end{align}
Thus, we deduce the following properties of anomalous dimensions
\begin{align}
\label{eq:currents-check}
\gamma_{\text{ns}}(1) = 0, && \begin{pmatrix} 1 && 2
\end{pmatrix}\begin{pmatrix}
\gamma_{qq}(2) & \gamma_{q\sigma}(2) \\
\gamma_{\sigma q}(2) & \gamma_{\sigma\sigma}(2)
\end{pmatrix} = 0,
\end{align}
which can be effectively used as a check of the obtained results. One can easily check that the obtained results~\eqref{eq:ns-results} and~\eqref{eq:singlet-results} satisfy~\eqref{eq:currents-check}.  
\subsection{$1/N$ expansion \label{subsec:large-N}}

The Gross-Neveu-Yukawa model~\eqref{eq:gny-action-def} is critically equivalent to the Gross-Neveu model~\cite{Gross:1974jv}. The critical behavior of the latter can be conveniently studied in the framework of the $1/N$ expansion for arbitrary $2 < d < 4$. This gives the possibility to study the behavior of critical exponents as analytic functions of $d$, see e.g.~\cite{Vasiliev:1992wr, Gracey:1992cp, Gracey:1993kc, Vasiliev:1993pi}. In particular, this means that expanding the results of the $1/N$ expansion in $\epsilon$ assuming $d = 4 - 2\epsilon$, we should reproduce the large-$N$ asymptotics of the results at the IR-stable critical point of the model~\eqref{eq:gny-action-def}.

Anomalous dimensions of the flavor non-singlet operator~\eqref{eq:nonsinglet-operator-def} were calculated to $1/N$ accuracy in~\cite{Muta:1976js} and have the form
\begin{align}
  \label{eq:large-N-non-singlet}
  \gamma^{1/N}_{\text{ns}}(s) = \dfrac{\eta_1}{\overline{N}}\left(1 - \dfrac{\mu(\mu - 1)}{(s + \mu - 1)(s + \mu - 2)}\right) + O\left(1/\overline{N}^2\right),
\end{align}
where $\mu = d/2$, $\overline{N} = N\cdot\operatorname{tr}\mathds{1}$ with $\mathds{1}$ being the identity operator in the Dirac space, and
\begin{equation}
  \eta_1 = - \dfrac{2\Gamma(2\mu - 1)}{\Gamma(\mu + 1)\Gamma(\mu)\Gamma(\mu - 1)\Gamma(1 - \mu)}.
\end{equation}
The result for the anomalous dimension~\eqref{eq:large-N-non-singlet} is also available in the $1/N^2$ approximation~\cite{Manashov:2016uam}, but we do not provide it here for brevity. Expanding~\eqref{eq:large-N-non-singlet} as a series in $\epsilon$ assuming $d = 4 - 2\epsilon$,~\footnote{In the expansion we use $\operatorname{tr}\mathds{1} = 4$.} we find perfect agreement with the results~\eqref{eq:ns-results}, namely
\begin{align}
  \label{eq:nons-inglet-large-N-check}
  \gamma^{1/N}_{\text{ns}}(s) = \gamma^*_{\text{ns}}(s) &= \Bigg\{\Bigg(\dfrac{1}{2} - \dfrac{1}{s(s + 1)}\Bigg)\epsilon - \Bigg(\dfrac{2s + 1}{s^2(s + 1)^2}  - \dfrac{3}{s(s + 1)} + \dfrac{3}{4}\Bigg)\epsilon^2 \nonumber \\  & \quad - \Bigg(\dfrac{1}{s^3(s+ 1)^3} - \dfrac{6s}{s^2(s + 1)^2} + \dfrac{2}{s(s + 1)} + \dfrac{3}{8}\Bigg)\epsilon^3 + O(\epsilon^4)\Bigg\}\dfrac{1}{N} \nonumber \\  &\quad + \Bigg\{\Bigg(\dfrac{3}{2s(s + 1)} - \dfrac{3}{4}\Bigg)\epsilon - \Bigg(\dfrac{3S_1}{2s(s + 1)} + \dfrac{2s + 1}{2s^3(s + 1)^3} - \dfrac{7s + 6}{2s^2(s + 1)^2} \nonumber \\ &\quad + \dfrac{17}{2s(s + 1)} - \dfrac{59}{16} + (-1)^s\left(\dfrac{1}{2s^2(s + 1)^2} - \dfrac{1}{2s(s + 1)}\right)\Bigg)\epsilon^2 - \Bigg(\dfrac{3S_2}{2s(s + 1)} \nonumber \\ & \quad + \left(\dfrac{3 + 6s}{2s(s + 1)} - \dfrac{25}{4}\right)\dfrac{S_1}{s(s + 1)} + \dfrac{3}{2s^4(s + 1)^4} - \dfrac{11s + 2}{s^3(s + 1)^3} + \dfrac{77s + 41}{4s^2(s + 1)^2} \nonumber \\ &\quad - \dfrac{27}{2s(s + 1)} + \dfrac{71}{32} + (-1)^s\Bigg(\left(\dfrac{1}{s(s + 1)} - 1\right)\dfrac{S_1}{s(s + 1)} + \dfrac{4s + 1}{2s^3(s + 1)^3} \nonumber \\ &\quad - \dfrac{s + 3}{s^2(s + 1)^2} + \dfrac{2}{s(s + 1)}\Bigg)\Bigg)\epsilon^3 + O(\epsilon^4)\Bigg\}\dfrac{1}{N^2} + O\big(1/N^3\big),
\end{align}
where $\gamma^*_{\text{ns}}(s) = \gamma_{\text{ns}}(s)\big|_{u = u^*, \lambda = \lambda^*}$ and the critical values of the couplings are taken from~\eqref{eq:couplings-critical}. One can see that even in the double series expansion ($\epsilon \to 0$, $N \to \infty$), the resulting expression~\eqref{eq:nons-inglet-large-N-check} is quite non-trivial, which serves as a comprehensive check.

In the singlet sector, the situation is more involved. The canonical scaling dimension of the field $\sigma(x)$ differs between the $\epsilon$ and $1/N$ expansions ($\Delta^{1/N}_{\sigma,0} = 1$ vs $\Delta^{\text{GNY}}_{\sigma, 0} = d/2 - 1$), which prevents the operators~\eqref{eq:singlet-operator-def} from mixing, and we therefore have two distinct anomalous dimensions~\cite{Muta:1976js,Giombi:2017rhm}
\begin{align}
  \gamma^{1/N}_{q}(s) &= \dfrac{\eta_1}{\overline{N}}\left(1 - \dfrac{\mu(\mu - 1)}{(s + \mu - 1)(s + \mu - 2)}\left(1 + \dfrac{\Gamma(2\mu - 1)\Gamma(s + 1)}{\Gamma(2\mu - 3  +s)(\mu - 1)}\right)\right) + O\big(1/\overline{N}^2\big), \\
  \gamma^{1/N}_{\sigma}(s) &= -\dfrac{\eta_1}{\overline{N}}\dfrac{2(2\mu - 1)}{\mu - 1}\left(1  - \dfrac{\mu}{2\mu - 1}\dfrac{\Gamma(\mu)\Gamma(s + 2 - \mu)}{\Gamma(\mu + s)\Gamma(3 - \mu)}\right) + O\big(1/\overline{N}^2\big).
\end{align}
The anomalous dimension of the fermion singlet operator is also known at order $1/N^2$~\cite{Manashov:2016uam}. Due to the critical equivalence, these anomalous dimensions should be associated with the eigenvalues of the singlet anomalous dimension matrix. Indeed, expanding the corresponding results, we find perfect agreement
\begin{align}
  \gamma_q^{1/N}(s) = \gamma^*_{-}(s) &= \Bigg\{\Bigg(\dfrac{1}{2} -\dfrac{3}{s(s + 1)}\Bigg)\epsilon - \Bigg(\dfrac{4S_1}{s(s + 1)} + \dfrac{6s + 3}{s^2(s + 1)^2} - \dfrac{13}{s(s + 1)} + \dfrac{3}{4}\Bigg)\epsilon^2 \nonumber \\
&\quad - \Bigg(\dfrac{4S_2}{s(s + 1)} + \dfrac{4S_1^2}{s(s + 1)} + \left(\dfrac{8s + 4}{s(s + 1)} - 20\right)\dfrac{S_1}{s(s + 1)} \nonumber \\
&\quad + \dfrac{3}{s^3(s + 1)^3} - \dfrac{26s + 4}{s^2(s + 1)^2} + \dfrac{18}{s(s + 1)} + \dfrac{3}{8}\Bigg)\epsilon^3\Bigg\}\dfrac{1}{N} \nonumber \\
&\quad + \Bigg\{\Bigg(\dfrac{3}{s^2(s + 1)^2} + \dfrac{4}{s(s + 1)} - \dfrac{3}{4}\Bigg)\epsilon + \Bigg(\left(\dfrac{10}{s(s + 1)} + \dfrac{9}{2}\right)\dfrac{S_1}{s(s + 1)} + \dfrac{6s - 3}{2s^3(s + 1)^3} \nonumber \\&\quad  + \dfrac{19s - 35}{2s^2(s + 1)^2} - \dfrac{127}{4s(s + 1)} + \dfrac{59}{16}\Bigg)\epsilon^2 + \Bigg(\left(\dfrac{4}{s(s + 1)} + \dfrac{11}{2}\right)\dfrac{S_2}{s(s + 1)} + \dfrac{4S_{-2}}{s^2(s + 1)^2} \nonumber \\&\quad + \left(\dfrac{18}{s(s + 1)} + 7\right)\dfrac{S_1^2}{s(s + 1)}  + \left(\dfrac{16s - 2}{s^2(s + 1)^2} + \dfrac{22s - 139}{2s(s + 1)} - \dfrac{187}{4}\right)\dfrac{S_1}{s(s + 1)} \nonumber \\ &\quad - \dfrac{36s + 21}{2s^4(s + 1)^4} - \dfrac{22s - 31}{2s^3(s + 1)^3} - \dfrac{289s - 149}{4s^2(s + 1)^2} + \dfrac{661}{8s(s + 1)} - \dfrac{71}{32}\Bigg)\epsilon^3 \nonumber \\&\quad + O(\epsilon^4)\Bigg\}\dfrac{1}{N^2} + O\big(1/N^3\big),
\end{align} 
and
\begin{align}
  \gamma_\sigma^{1/N}(s) = \gamma^*_{+}(s) &= \Bigg\{\Bigg(\dfrac{2}{s(s + 1)} - 3\Bigg)\epsilon + \Bigg(\dfrac{4S_1}{s(s + 1)} - \dfrac{2}{s^2(s + 1)^2} - \dfrac{4}{s(s + 1)} + \dfrac{7}{2}\Bigg)\epsilon^2 \nonumber \\
&\quad + \Bigg(\dfrac{4S_1^2}{s(s + 1)} - \left(\dfrac{4}{s(s + 1)} + 8\right)\dfrac{S_1}{s(s + 1)} + \dfrac{2}{s^3(s + 1)^3} + \dfrac{6}{s^2(s + 1)^2} + \dfrac{11}{4}\Bigg)\epsilon^3  \nonumber \\ &\quad + O(\epsilon^4)\Bigg\}\dfrac{1}{N}  + O\big(1/N^2\big),
\end{align}
where $\gamma^{*}_{\pm}(s) = \gamma_{\pm}(s)\big|_{u = u^*, \lambda = \lambda^*}$ and the eigenvalues, $\gamma_{\pm}(s)$, are defined in~\eqref{eq:eigenvalues-definition}.

\subsection{Reciprocity \label{subsec:reciprocity}}

An important property of the anomalous dimensions of the leading-twist operators is the 
generalized Gribov-Lipatov reciprocity~\cite{Dokshitzer:2005bf, Dokshitzer:2006nm, 
Basso:2006nk}. This property constrains the large-spin asymptotics of the anomalous 
dimension. 
As a rule, in the literature, the reciprocity 
property is studied in models with one coupling, which allows one to extend the properties 
of the critical values of anomalous dimensions to arbitrary coupling using the 
$\epsilon$-independence of the coefficients in the $\overline{\text{MS}}$ scheme. The 
model~\eqref{eq:gny-action-def} has two couplings, $u$ and $\lambda$, which makes such a 
transition impossible, and reciprocity can be studied only at the critical 
point~\eqref{eq:betas-zero} (see~\cite{Alday:2015eya} for the derivation of reciprocity in the CFT setting).

For the critical value of the anomalous dimension of a given leading-twist operator, 
$\gamma^*(s)$, we introduce the change of variables
\begin{equation}
  \label{eq:reciprocity-change-def}
  \gamma^*(s) = f\left(s - \epsilon + \dfrac{1}{2}\gamma^*(s)\right),
\end{equation}
where $j \equiv s - \epsilon + \frac{1}{2}\gamma^*(s)$ is the conformal spin. We call the 
resulting function $f(j)$ reciprocity-respecting (RR) if its asymptotics in the limit 
$j \to \infty$ is invariant under the parity transform $j \mapsto -j - 1$. The 
corresponding anomalous dimension $\gamma^*(s)$ then satisfies the generalized 
Gribov-Lipatov reciprocity property.

In this section we demonstrate that all obtained results in Section~\ref{sec:res} satisfy the reciprocity 
property. In practice, rather than working with the large-$j$ asymptotics, it is more 
convenient to check the following equivalent properties:
\begin{itemize}
  \item All rational functions entering $f(j)$ are explicitly invariant under 
  $j \mapsto -j - 1$.
  \item All harmonic sums are organized in reciprocity-respecting (RR) 
  combinations~\cite{Beccaria:2007bb, Beccaria:2009vt, Beccaria:2010tb}.~\footnote{Note 
  also the alternative construction of binomial harmonic sums~\cite{Weinzierl:2004bn, 
  Ablinger:2014bra}, whose asymptotic expansion also satisfies the parity transform.}
\end{itemize}
For the purposes of this paper we need a rather modest set of such sums
\begin{align}
  \label{eq:rr-harmonic-sums}
  \Omega_1 = S_1, && \Omega_{-2} = S_{-2}, && \Omega_{3} = S_3, && \Omega_{1, -2} = S_{1, -2} 
  - \dfrac{1}{2}S_{-3},
\end{align} 
the general construction can be found e.g. in~\cite{Beccaria:2009vt}.

\paragraph{Non-singlet case:} We begin our check with the flavor non-singlet anomalous 
dimensions~\eqref{eq:ns-results}. We note that the change of 
variables~\eqref{eq:reciprocity-change-def} in principle requires analytic continuation 
from $s \in \mathbb{Z}_{> 0}$ to $s \in \mathbb{C}\backslash\{0, -1, -2, \ldots\}$. In 
the non-singlet case, such a continuation should be done independently for the odd and even 
values of $s$, yielding two different continued functions: $\gamma^{+}_{\text{ns}}(s)$ 
defined as the analytic continuation from $s \in 2\mathbb{Z}_{> 0}$ and 
$\gamma^{-}_{\text{ns}}(s)$ continued from $s \in 2\mathbb{Z}_{>0} + 1$. As a result, we 
consider two distinct RR functions, $f_{\text{ns}}^{+}(j)$ and $f_{\text{ns}}^{-}(j)$.

The change of variables~\eqref{eq:reciprocity-change-def} is convenient to perform order 
by order in perturbation theory
\begin{subequations}
\label{eq:reciprocity-non-singlet-pert}
\begin{align}
  \big(f_{\text{ns}}^{\pm}\big)^{(1)}(j) &= \big(\gamma^{\pm, *}_{\text{ns}}\big)^{(1)}(j), \\
  \big(f_{\text{ns}}^{\pm}\big)^{(2)}(j) &= \big(\gamma^{\pm, *}_{\text{ns}}\big)^{(2)}(j) - 
  \partial_j\big(f_{\text{ns}}^{\pm}\big)^{(1)}(j)\left(\dfrac{1}{2}
  \big(\gamma^{\pm, *}_{\text{ns}}\big)^{(1)}(j) - 1\right), \\
  \big(f_{\text{ns}}^{\pm}\big)^{(3)}(j) &= \big(\gamma^{\pm, *}_{\text{ns}}\big)^{(3)}(j) - 
  \partial_j\big(f_{\text{ns}}^{\pm}\big)^{(2)}(j)\left(\dfrac{1}{2}
  \big(\gamma^{\pm, *}_{\text{ns}}\big)^{(1)}(j) - 1\right) - 
  \partial_j\big(f_{\text{ns}}^{\pm}\big)^{(1)}(j)\dfrac{1}{2}
  \big(\gamma^{\pm, *}_{\text{ns}}\big)^{(2)}(j) \nonumber \\ 
  &\quad -\dfrac{1}{2}\partial_j^2\big(f_{\text{ns}}^{\pm}\big)^{(1)}(j)\left(\dfrac{1}{2}
  \big(\gamma^{\pm, *}_{\text{ns}}\big)^{(1)}(j) - 1\right)^2, 
\end{align}
\end{subequations}
where $f_{\text{ns}}^{\pm}(j) = \big(f_{\text{ns}}^{\pm}\big)^{(1)}\epsilon + \big(f_{\text{ns}}^{\pm}\big)^{(2)}\epsilon^2 + 
\ldots$, etc. Substituting the obtained results, we find
\begin{align}
  f_{\text{ns}}^{+}(j) &= u_1\left\{1 - \dfrac{2}{j(j + 1)}\right\}\epsilon + \Bigg\{u_2\left(1 - 
  \dfrac{2}{j(j + 1)}\right) + u_1^2\Bigg(-\dfrac{6\Omega_1}{j(j + 1)} + 
  \dfrac{3}{j^2(j + 1)^2} + \dfrac{12N + 16}{j(j + 1)} \nonumber \\ &\quad  - 
  \dfrac{12N + 1}{4}\Bigg)\Bigg\}\epsilon^2  + \Bigg\{u_3\left(1 - \dfrac{2}{j(j + 
  1)}\right) + u_1u_2\Bigg(- \dfrac{12\Omega_1}{j(j + 1)}  + \dfrac{6}{j^2(j + 1)^2} + 
  \dfrac{24N + 32}{j(j + 1)} \nonumber \\ &\quad  - \dfrac{12N + 1}{2}\Bigg) + 
  \lambda_1u_1^2\Bigg(\dfrac{8\Omega_1}{j(j + 1)} - \dfrac{18}{j(j + 1)} + 2\Bigg) + 
  \dfrac{\lambda_1^2u_1}{24}\left(\dfrac{17}{j(j + 1)} - \dfrac{11}{2}\right) \nonumber \\ 
  &\quad - u_1^3\Bigg(\left(1 + \dfrac{1}{j(j + 1)}\right)\dfrac{8\Omega_{1,- 2} + 
  4\Omega_3}{j(j + 1)} + \dfrac{9\Omega_1^2}{j(j + 1)} - \left(\dfrac{8}{j(j + 1)} + 
  20\right)\dfrac{\Omega_{-2}}{j^2(j + 1)^2} \nonumber \\&\quad + \Bigg(\dfrac{8N - 
  13}{j(j + 1)} - \dfrac{116N + 81}{2}\Bigg)\dfrac{\Omega_1}{j(j + 1)}- 
  \left(\dfrac{12}{j^2(j + 1)^2} + 3\right)\zeta_3  + \Bigg(\dfrac{6}{j(j + 1)} + 12N 
  \nonumber \\ &\quad + 15\Bigg)\dfrac{\zeta_2}{j(j + 1)}  + \dfrac{13}{2j^3(j+ 1)^3} + 
  \dfrac{132N + 155}{4j^2(j + 1)^2} + \dfrac{128N^2 + 1104N + 275}{8j(j + 1)}\nonumber 
  \\&\quad + \dfrac{48N^2 - 188N +15}{16}\Bigg)\Bigg\}\epsilon^3 + O(\epsilon^4), 
\end{align}
\begin{align}
  f_{\text{ns}}^{-}(j) &= u_1\left\{1 - \dfrac{2}{j(j + 1)}\right\}\epsilon + \Bigg\{u_2\left(1 - 
  \dfrac{2}{j(j + 1)}\right) + u_1^2\Bigg(-\dfrac{6\Omega_1}{j(j + 1)} + 
  \dfrac{7}{j^2(j + 1)^2} + \dfrac{12N + 12}{j(j + 1)} \nonumber \\ &\quad  - 
  \dfrac{12N + 1}{4}\Bigg)\Bigg\}\epsilon^2  + \Bigg\{u_3\left(1 - \dfrac{2}{j(j + 
  1)}\right) + u_1u_2\Bigg(- \dfrac{12\Omega_1}{j(j + 1)} + \dfrac{14}{j^2(j + 1)^2} + 
  \dfrac{24N + 24}{j(j + 1)} \nonumber \\
&\quad  - \dfrac{12N + 1}{2}\Bigg)+ \lambda_1u_1^2\Bigg(\dfrac{8\Omega_1}{j(j + 1)} - 
\dfrac{18}{j(j + 1)} + 2\Bigg) + \dfrac{\lambda_1^2u_1}{24}\left(\dfrac{17}{j(j + 1)} - 
\dfrac{11}{2}\right) \nonumber \\
&\quad - u_1^3\Bigg(\left(-1 + \dfrac{3}{j(j + 1)}\right)\dfrac{8\Omega_{1,- 2} + 
4\Omega_3}{j(j + 1)} + \dfrac{9\Omega_1^2}{j(j + 1)} + \left(\dfrac{4}{j(j + 1)} + 
16\right)\dfrac{\Omega_{-2}}{j(j + 1)} \nonumber \\
&\quad - \Bigg(\dfrac{8N + 17}{j(j + 1)} + \dfrac{84N + 57}{2}\Bigg)\dfrac{\Omega_1}{j(j 
+ 1)} - \left(\dfrac{36}{j^2(j + 1)^2} - \dfrac{24}{j(j + 1)} + 3\right)\zeta_3 \nonumber 
\\
&\quad + \Bigg(\dfrac{6}{j(j + 1)} + 12N + 15\Bigg)\dfrac{\zeta_2}{j(j + 1)} + 
\dfrac{4}{j^4(j + 1)^4} + \dfrac{16N + 69}{2j^3(j + 1)^3} \nonumber \\
&\quad + \dfrac{292N + 483}{4j^2(j + 1)^2} + \dfrac{128N^2 + 848N - 77}{8j(j + 1)} + 
\dfrac{48N^2 - 188N + 15}{16}\Bigg)\Bigg\}\epsilon^3 + O(\epsilon^4). 
\end{align}
As is easy to notice, all the rational functions are expressed through the combination 
$j(j + 1)$, while the harmonic sums appear only in the RR combinations~\eqref{eq:rr-harmonic-sums}, 
so the non-singlet anomalous dimensions possess the generalized Gribov-Lipatov reciprocity 
property. It is important to note that this property holds for the $\gamma_{\text{ns}}^+(s)$ and $\gamma_{\text{ns}}^{-}(s)$ independently. 

Another property of the non-singlet anomalous dimension, which might be useful in the 
process of analytical reconstruction discussed in Section~\ref{subsec:method}, is that one can 
consider the reciprocity property in the limit $\lambda \to 0$. In such a limit, the 
theory~\eqref{eq:gny-action-def} has only one coupling constant, which allows one to use 
the simpler version of the transformation~\eqref{eq:reciprocity-change-def}, namely
\begin{equation}
\label{eq:reciprocity-reduced}
\gamma(s)\Big|_{\lambda \to 0} = f\left(s + \bar{\beta}_u(u, 0) + \dfrac{1}{2}\gamma(s)
\Big|_{\lambda \to 0}\right),
\end{equation}
where $\beta_u(u, \lambda) = -2u(\epsilon + \bar{\beta}(u, \lambda))$, 
see~\eqref{eq:beta-u}. Note that the anomalous dimension $\gamma(s)\big|_{\lambda \to 0}$ 
here is considered at arbitrary coupling $u$. The RR form of the corresponding function 
$f(j)$ is then justified by the $\epsilon$-independence of the perturbative coefficients 
in the $\overline{\text{MS}}$ scheme, which in the case of one coupling allows one to 
perform the change $\epsilon \mapsto -\bar{\beta}(u)$ in~\eqref{eq:reciprocity-change-def}. 
From the technical point of view, working with~\eqref{eq:reciprocity-reduced} is much 
simpler, as it allows one to bypass the cumbersome expressions for the critical 
couplings~\eqref{eq:couplings-critical}. Here we do not present the results for the 
reduced functions $f(j)$ corresponding to $\gamma^{\pm}_{\text{ns}}(s)\big|_{\lambda \to 
0}$, while noting that their RR form has been verified.

\paragraph{Singlet case:}  When we consider the anomalous dimensions in the flavor singlet sector, we have to consider 
two important changes. The first, minor change is related to the fact that the anomalous 
dimension matrix is originally defined only for the even values of spin, $s \in 2\mathbb{Z}_{> 0}$, 
so in what follows we consider only functions obtained by analytic continuation from even 
values. The major change is, however, that one can immediately find that the separate 
entries of the anomalous dimension matrix do not possess the reciprocity property.

This problem is well known (see e.g.~\cite[Section 2.2]{Basso:2006nk} or~\cite{Chen:2020uvt} 
for a more recent discussion) and the solution consists in considering the eigenvalues of 
the anomalous dimension matrix.~\footnote{This fact also fits well with the line of 
work~\cite{Alday:2015eya}, as precisely the eigenvalues of the dilatation operator take 
part in the spectrum of the critical CFT.} Thus, we define two RR functions
\begin{equation}
  \gamma^*_{\pm}(s) = f_{\pm}\left(s - \epsilon + \dfrac{1}{2}\gamma^*_{\pm}(s)\right),
\end{equation}
where $\gamma^*_{\pm}(s) = \gamma_{\pm}(s)\big|_{u = u^*, \lambda = \lambda^*}$ are the 
critical values of the eigenvalues of the singlet anomalous dimension matrix. In practice, 
it is technically complicated to work with the eigenvalues $f_{\pm}(s)$ directly, as the 
process of diagonalization introduces square roots. It is, however, easy to show that 
instead of the eigenvalues we can equivalently consider any set of independent matrix 
invariants. For the purposes of the present paper we use the trace and the determinant, 
which can be expressed as the sum and product of the functions $f_{\pm}(j)$.

To derive the corresponding perturbative expansions we use the convenient notation
\begin{align}
  T(s) = \operatorname{tr}\widehat{\gamma}^*(s), && D(s) = \operatorname{det}\widehat{\gamma}^*(s).
\end{align}
The singlet analog of the change of variables~\eqref{eq:reciprocity-non-singlet-pert} then 
reads
\begin{subequations}
  \begin{align}
    \left(f_{+} + f_{-}\right)^{(1)}(j) &=  T^{(1)}(j),\\
    \left(f_{+} + f_{-}\right)^{(2)}(j) &=  T^{(2)}(j) - \partial_jT^{(1)}(j)\left(\dfrac{1}{2}T^{(1)}(j) - 1\right) + \dfrac{1}{2}\partial_j D^{(2)}(j), \\
    \left(f_{+} + f_{-}\right)^{(3)}(j) &= T^{(3)}(j) - \dfrac{1}{2}\partial_j T^{(1)}(j)\left(T^{(2)}(j) + \dfrac{1}{2}\partial_j D^{(2)}(j)\right) \nonumber  \\
    &\quad + \left(\dfrac{1}{2}T^{(1)}(j) - 1\right)\left(\dfrac{1}{2}\left(\partial_j T^{(1)}(j)\right)^2 - \partial_j T^{(2)}(j)\right) \nonumber \\
    &\quad + \dfrac{1}{2}\partial_j^2 T^{(1)}(j)\left(\left(\dfrac{1}{2}T^{(1)}(j) - 1\right)^2 - \dfrac{1}{4}D^{(2)}(j)\right) \nonumber \\
    &\quad - \dfrac{1}{2}\partial_j^2 D^{(2)}(j)\left(\dfrac{1}{4}T^{(1)}(j) - 1\right) + \dfrac{1}{2}\partial_j D^{(3)}(j),
  \end{align}
\end{subequations}
\begin{subequations}
  \begin{align}
    \left(f_{+}\cdot f_{-}\right)^{(2)}(j) &= D^{(2)}(j)\\
    \left(f_{+}\cdot f_{-}\right)^{(3)}(j) &= D^{(3)}(j) - \dfrac{1}{2}\partial_j T^{(1)}(j)D^{(2)}(j) + \partial_jD^{(2)}(j)\\
    \left(f_{+}\cdot f_{-}\right)^{(4)}(j) &= D^{(4)}(j) - \dfrac{1}{2}\partial_j T^{(1)}(j)\left(D^{(3)}(j) + \partial_j D^{(2)}(j)\right) + \partial_j D^{(3)}(j) \nonumber \\
    &\quad + D^{(2)}(j)\left(\dfrac{1}{4}\left(\partial_j T^{(1)}(j)\right)^2 - \dfrac{1}{2}\partial_j T^{(2)}(j)\right) \nonumber \\
    &\quad + \dfrac{1}{2}\partial_j^2 D^{(2)}(j)\left(1 - \dfrac{1}{4}D^{(2)}(j)\right) + \dfrac{1}{4}\partial_j^2 T^{(1)}(j)\, D^{(2)}(j)\left(\dfrac{1}{2}T^{(1)}(j) - 1\right),
  \end{align}
\end{subequations}
where
\begin{align}
  T(s) = T^{(1)}(s)\epsilon + T^{(2)}(s)\epsilon^2 + O(\epsilon^3), && D(s) = D^{(2)}(s)\epsilon^2 + D^{(3)}(s)\epsilon^3 + O(\epsilon^4).
\end{align}
Using these formulas we obtain
\begin{align}
  \left(f_{+} + f_{-}\right)(j) &= u_1\Bigg\{1 + 4N - \dfrac{2}{j(j + 1)}\Bigg\}\epsilon + \Bigg\{u_2\Bigg(1 + 4N - \dfrac{2}{j(j + 1)}\Bigg) + \lambda_1^2\Bigg(\dfrac{1}{6} - \dfrac{1}{j(j + 1)}\Bigg) \nonumber \\ &\quad -u_1^2\Bigg(\dfrac{6\Omega_1}{j(j+ 1)}   + \dfrac{8N - 3}{j^2(j + 1)^2} - \dfrac{36N + 16}{j(j + 1)}  + \dfrac{52N + 1}{4}\Bigg)\Bigg\}\epsilon^2 + \Bigg\{u_3\Bigg(1 + 4N \nonumber \\
  &\quad  - \dfrac{2}{j(j + 1)}\Bigg) + \lambda_1\lambda_2\Bigg(\dfrac{1}{3} - \dfrac{2}{j(j + 1)}\Bigg) - u_1u_2\Bigg(\dfrac{12\Omega_1}{j(j + 1)} + \dfrac{16N - 6}{j^2(j + 1)^2} - \dfrac{72N + 32}{j(j + 1)}\nonumber \\
  &\quad + 26N + \dfrac{1}{2}\Bigg)  - \lambda_1^2u_1\Bigg(\dfrac{4N\Omega_1}{j(j + 1)} - \dfrac{2N}{j^2(j + 1)^2} - \dfrac{288N + 17}{24j(j + 1)} + \dfrac{60N + 11}{48}\Bigg)  \nonumber \\
  &\quad  + \lambda_1u_1^2\Bigg(\dfrac{(8 + 48N)\Omega_1}{j(j + 1)}- \dfrac{64N + 18}{j(j + 1)} + 2 + 10N\Bigg) - \lambda_1^3\Bigg(\dfrac{2\Omega_1}{j(j + 1)} - \dfrac{5}{j(j + 1)} + \dfrac{1}{8}\Bigg)  \nonumber \\
  &\quad  - u_1^3\Bigg(\left(\dfrac{4}{j(j + 1)} + 16N + 4\right)\dfrac{2\Omega_{1,- 2} + \Omega_3}{j(j + 1)} - \Bigg(\dfrac{8}{j^2(j + 1)^2} + \dfrac{32N + 20}{j(j + 1)}  \nonumber \\
  &\quad - 32N\Bigg)\dfrac{\Omega_{-2}}{j(j + 1)} + \dfrac{9\Omega_1^2}{j(j + 1)} + \Bigg(\dfrac{32N^2 + 48N - 13}{j(j + 1)} - \dfrac{192N^2 + 292N + 81}{2}\Bigg)\dfrac{\Omega_1}{j(j + 1)} \nonumber \\
  &\quad  - \left(\dfrac{12}{j^2(j + 1)^2} + \dfrac{48N}{j(j + 1)} + 3(4N + 1)\right)\zeta_3 + \Bigg(\dfrac{6}{j(j + 1)} + 32N^2 + 52N  \nonumber \\
  &\quad  + 15\Bigg)\dfrac{\zeta_2}{j(j + 1)} - \dfrac{32N^2 + 40N - 13}{2j^3(j + 1)^3} - \dfrac{192N^2 + 348N - 155}{4j^2(j + 1)^2}  \nonumber \\
  &\quad  + \dfrac{1152N^2 + 3600N + 275}{8j(j + 1)} - \dfrac{752N^2 + 272N - 15}{16}\Bigg)\Bigg\}\epsilon^3 + O(\epsilon^4),
\end{align}
\begin{align}
  \left(f_{+}\cdot f_{-}\right)(j) &= 4Nu_1^2\Bigg\{1 - \dfrac{6}{j(j + 1)}\Bigg\}\epsilon^2 + \Bigg\{8Nu_2u_1\Bigg(1 - \dfrac{6}{j(j + 1)}\Bigg) + \lambda_1^2u_1\Bigg(\dfrac{2}{j^2(j + 1)^2} - \dfrac{4}{3j(j + 1)}\nonumber \\&\quad + \dfrac{1}{6}\Bigg) - Nu_1^3\Bigg(\dfrac{64N + 104}{j(j + 1)}\Omega_1 + \dfrac{60}{j^2(j + 1)^2} - \dfrac{208N + 292}{j(j + 1)} + 12N + 11\Bigg)\Bigg\}\epsilon^3 \nonumber \\ &\quad  + \Bigg\{4N(2u_3u_1 + u_2^2)\Bigg(1 - \dfrac{6}{j(j + 1)}\Bigg) + \lambda_1^2u_2\Bigg(\dfrac{2}{j^2(j + 1)^2} - \dfrac{4}{3j(j + 1)} + \dfrac{1}{6}\Bigg) \nonumber \\
  &\quad + \lambda_1\lambda_2u_1\Bigg(\dfrac{4}{j^2(j + 1)^2} - \dfrac{8}{3j(j + 1)} + \dfrac{1}{3}\Bigg) \nonumber \\
  &\quad - Nu_2u_1^2\Bigg(\dfrac{(192N + 312)\Omega_1}{j(j + 1)} + \dfrac{180}{j^2(j + 1)^2} - \dfrac{624N + 876}{j(j + 1)} + 36N + 33\Bigg) \nonumber \\
  &\quad - N\lambda_1u_1^3\Bigg(\left(\dfrac{96}{j(j + 1)} - 144\right)\dfrac{\Omega_1}{j(j + 1)} - \dfrac{160}{j^2(j + 1)^2} + \dfrac{380}{j(j + 1)} - 18\Bigg) \nonumber \\
  &\quad + \lambda_1^2u_1^2\Bigg(\left(\dfrac{8N + 6}{j(j + 1)} - \dfrac{20N + 3}{3}\right)\dfrac{\Omega_1}{j(j + 1)} - \dfrac{4N + 3}{j^3(j + 1)^3} - \dfrac{52N + 31}{2j^2(j + 1)^2} \nonumber \\
  &\quad + \dfrac{388N + 35}{12j(j + 1)} - \dfrac{64N + 1}{24}\Bigg) \nonumber \\
  &\quad + \lambda_1^3u_1\Bigg(\left(\dfrac{4}{j(j + 1)} - 2\right)\dfrac{\Omega_1}{j(j + 1)} - \dfrac{10}{j^2(j + 1)^2} + \dfrac{21}{4j(j + 1)} - \dfrac{1}{8}\Bigg) \nonumber \\
  &\quad - Nu_1^4\Bigg(\left(\dfrac{48}{j(j + 1)} + 32\right)\dfrac{2\Omega_{1,- 2} + \Omega_3}{j(j + 1)} - \left(\dfrac{96}{j^2(j + 1)^2} + \dfrac{128N + 592}{j(j + 1)} - 32\right)\dfrac{\Omega_{-2}}{j(j + 1)} \nonumber \\
  &\quad + \dfrac{(128N^2 + 320N + 236)\Omega_1^2}{j(j + 1)} + \Bigg(\dfrac{320N + 276}{j(j + 1)} - 640N^2 - 2200N - 1298\Bigg)\dfrac{\Omega_1}{j(j + 1)} \nonumber \\
  &\quad - \left(\dfrac{144}{j^2(j + 1)^2} - \dfrac{168}{j(j + 1)} + 24\right)\zeta_3 + \left(\dfrac{72}{j(j + 1)} + 128N^2 + 400N + 300\right)\dfrac{\zeta_2}{j(j + 1)} \nonumber \\
  &\quad - \dfrac{48N + 72}{j^4(j + 1)^4} - \dfrac{424N + 270}{j^3(j + 1)^3} - \dfrac{460N + 1069}{j^2(j + 1)^2} + \dfrac{576N^2 + 3396N + 1366}{j(j + 1)} \nonumber \\
  &\quad + 12N^2 - 127N - 4\Bigg)\Bigg\}\epsilon^4 + O(\epsilon^5).
\end{align}
Here we again see that both $f_{+} + f_{-}$ and $f_{+}\cdot f_{-}$ are expressed through 
the combination $j(j + 1)$ and the RR harmonic sums~\eqref{eq:rr-harmonic-sums}, which 
means that the expressions in~\eqref{eq:singlet-results} possess the generalized Gribov-Lipatov reciprocity 
property.
\subsection{Emergence of SUSY}
As a final consistency check, we can verify whether the general $s$-expressions we derived respect the SUSY Ward Identity \eqref{bare_ward} at the emergent SUSY points. To that purpose, rather than just the Gross-Neveu-Yukawa model, we need the general expressions for the generalised Lagrangian $\lgen$ from Appendix~\ref{app:lgen}.

As was established in \cite{Chakraborty:2026lna}, the generalised Lagrangian $\lgen$ has two emergent SUSY points corresponding to the 3D and 4D Wess-Zumino models. By the conventions used in this paper, substituting $\lambda=3u$ and $(N_s,N_f)=(1,1/4)$ and $(2,1/2)$ should make SUSY Ward identities emerge among the non-supersymmetric quantities of $\lgen$. For our singlet operators, this amounts to the following quantity being zero at all loop orders and at all moments $s$:~\footnote{This relation is similar to the famous Dokshitzer relation~\cite{Dokshitzer:1977sg}, which is related to the supersymmetry in $\mathcal{N} = 1$ SYM, see~\cite{Antoniadis:1981zv} for details.}
\begin{equation}
\mathcal{R}(s)\;\equiv\;
2\gamma_{qq}+2s\,\gamma_{\sigma q}
-\tfrac{2}{s}\,\gamma_{q\sigma}-2\gamma_{\sigma\sigma}\;=\;0 ,
\label{eq:susy}
\end{equation}
which is just the SUSY Ward Identity \eqref{bare_ward}. Our computations used the emergent SUSY condition for only the non-planar topology, and that too only at some fixed values of  $s$. We want to check that the full anomalous dimension respects the emergence of SUSY at any general value of $s$. 

For an appropriate analysis, the 3D extension of the anomalous dimensions needs to be accurately considered. This is taken care of in the results of Appendix~\ref{app:lgen}, the 3D enhancement is denoted by the symbol $\eta$, with $\eta=0$ for 4D results and $\eta=1$ for the 3D extension. It so turns out that the equation \eqref{eq:susy} is an identity in $s$ for
\begin{equation}\label{eq:emgsusypts}
(N_s,N_f)=\left(2,\tfrac12\right) \ \ \text{for}\ \eta=0 ,
\qquad\qquad
(N_s,N_f)=\left(1,\tfrac14\right) \ \ \text{for}\ \eta=1 ,
\end{equation}
which is precisely the emergence of SUSY as predicted by \cite{Chakraborty:2026lna}. In fact, the above result also provides further evidence that the 3D enhancement term is indispensable for the 3D extension of the Gross-Neveu-Yukawa result, which is significant for applications to the study of the chiral Ising model. This is because, without the enhancement, SUSY does not emerge at 3 loops for the $(N_s,N_f)=(1,1/4)$ point:
\begin{alignat}{2}
\eta=0:&\quad \mathcal{R}(s)\big|_{(1,1/4)}
 &&= \phantom{-}\frac{8\,(s^2+s-2)}{s^2(s+1)^2}\bigl(2S_{-2}+S_1\bigr) ,
\end{alignat}
which indicates that without the $\eta$-term it is not a truly 3D quantity.

Furthermore, it was observed and conjectured in \cite{Chakraborty:2026lna} that for the highest transcendental weights at every loop order, SUSY relations emerge simply from the equality of fermionic and bosonic degrees of freedom, rather than requiring the exact SUSY multiplet. That amounts to $\mathcal{R}(s)$ being proportional to a factor of $(N_s-4N_f)$ at those weights. The general $s$ results confirm that hypothesis in our case. With $C$ as in Eq.~\eqref{eq:Cdef} and
\begin{equation}
F_3\equiv S_3(s)-S_{-3}(s)+2S_{1,-2}(s) ,
\end{equation}
the entire weight-3 content of $\mathcal{R}$ is the single combination $F_3$,
carrying a linear factor $(N_s-4N_f)$:
\begin{equation}
\mathcal{R}\big|_{\text{wt-}3} = \frac{8 C (N_s-4N_f)}{s^2}\;F_3 ,
\qquad
\mathcal{R}\big|_{\zeta_3} = \frac{24 (s-1) C (N_s-4N_f)}{s^2} .
\end{equation}
Some lower weight-2 terms $S_1^2$, $S_2$ are also proportional to $(N_s-4N_f)$ in addition to the highest weights, refer to the table below:
\begin{center}
\begin{tabular}{lcc}
\hline
structure & divisible by $(N_s-4N_f)$ & divisible by $C$ \\
\hline
$S_3,\;S_{-3},\;S_{1,-2},\;\zeta_3$ & yes & yes \\
$S_1^2,\;S_2$                        & yes & no  \\
$S_{-2},\;S_1$, rational             & no  & no  \\
\hline
\end{tabular}
\end{center}
Also notice the divisibility by $C$ in the table above. The flavour factor $C$ is associated with the non-planar topology, which is the heaviest part of the computation. Hence, it is an indicator that the highest weight harmonic sums in the operator anomalous dimensions are almost entirely contributed by the non-planar diagrams, the diagrams that we optimised in Section~\ref{sec:opmitsusy}.\\[3pt]
The factor $(N_s-4N_f)$ organizes the two highest factors of $s(s+1)$ as well:
\begin{align}
\mathcal{R}\Big|_{1/(s(s+1))^{4}}
 &= -2\, N_s (N_s-4N_f)\Bigl[(24+20N_f)(s+1)-(11s+4)\,N_s\Bigr] ,
\\
\mathcal{R}\Big|_{1/(s(s+1))^{5}}
 &= -8\,(s+1)\, N_s^{2} (N_s-4N_f) .
\end{align}
These highest factors can be traced back to one-loop propagator insertions, and thus carry the $(N_s-4N_f)$ factor similar to one loop (see \cite{Chakraborty:2026lna}).

\section{The small-spin limit\label{sec:small-spin}}

\begin{figure}[h]
\centering
\includegraphics[width=0.6\columnwidth]{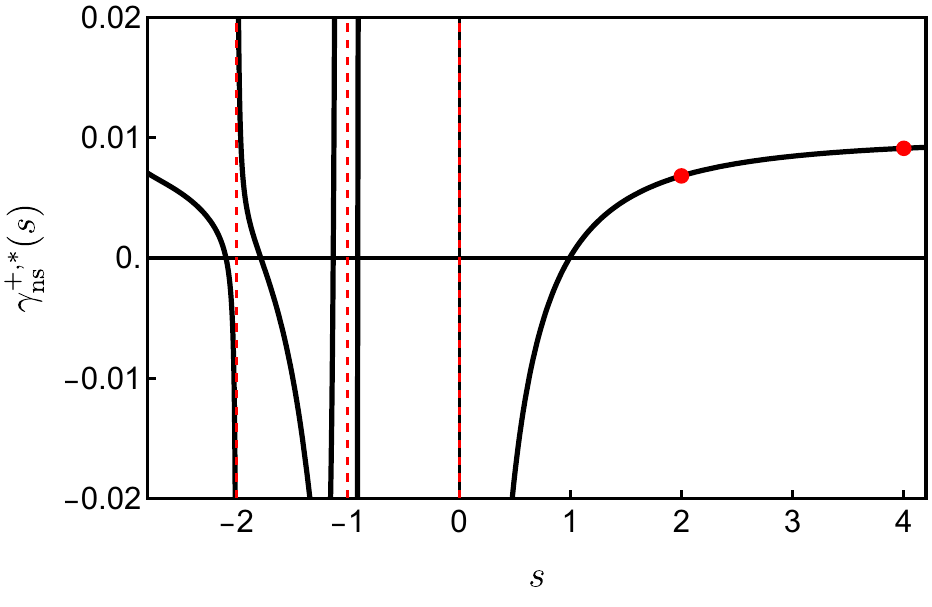}
\caption{Analytic continuation of the critical anomalous dimension 
$\gamma^{+,*}_{\text{ns}}(s)$. Red dashed lines represent the singularities arising in the 
region $s \le 0$, while red dots correspond to the anomalous dimensions of the local 
operators~\eqref{eq:nonsinglet-operator-def} for even values of $s$. The graph is made for 
$N = 1$, $\epsilon = 0.05$.}
\label{fig:ns-singularities}
\end{figure}

In Section~\ref{subsec:reciprocity} we discussed the remarkable properties of the obtained 
anomalous dimensions as functions of spin in the large-$s$ limit, which are related to the 
conformal symmetry arising at the critical point~\eqref{eq:couplings-critical}. In this section we discuss what constraints this symmetry puts on the small-$s$ limit. It 
is easy to see that the analytic continuations of the obtained anomalous 
dimensions behave as smooth analytic functions in the region 
$\operatorname{Re} s > 0$, while containing poles in $\operatorname{Re} s \le 0$ (see 
Fig.~\ref{fig:ns-singularities}). It turns out that the behavior of the critical anomalous 
dimensions around these singularities is not arbitrary, which allows us to perform a 
resummation of such singular contributions in the perturbative series. In what follows we 
revisit such a resummation procedure in the case $s \to 0$ (see~\cite{Manashov:2025kgf} 
for the two-loop case) and discuss its possible practical implementations.

\subsection{Conformal Regge trajectories \label{subsec:regge}}
  
The situation with the pole at $s \to 0$ strongly resembles the singularity arising in the 
twist-2 anomalous dimensions in the $\varphi^4$ theory, which is discussed in detail 
in~\cite{Caron-Huot:2022eqs}. Let us adapt their arguments to the case of the 
model~\eqref{eq:gny-action-def}. A convenient tool for this is the conformal Chew-Frautschi 
plot, which can be seen as the real slice of the plot $(s, \Delta^*(s))$, $s \in \mathbb{C}$. 
The analytic continuation of the scaling dimensions of leading-twist operators $\Delta^*(s)$ 
defines conformal Regge trajectories (see~\cite{Costa:2012cb} for details) on this plot. In 
the case of leading-twist operators we have the following four trajectories:
\begin{itemize}
  \item $\Delta^{+, *}_{\text{ns}}(s) = 2 - 2\epsilon + s + \gamma^{+, *}_{\text{ns}}(s)$ 
  defined by the non-singlet anomalous dimensions~\eqref{eq:ns-results}, continued from 
  $s \in 2\mathbb{Z}_{>0}$. 
  \item $\Delta^{-, *}_{\text{ns}}(s) = 2 - 2\epsilon + s + \gamma^{-, *}_{\text{ns}}(s)$ 
  defined by the non-singlet anomalous dimensions~\eqref{eq:ns-results}, continued from 
  $s \in 2\mathbb{Z}_{\ge 0} + 1$.
  \item $\Delta^*_{\pm}(s) = 2 - 2\epsilon + s + \gamma^*_{\pm}(s)$ --- two trajectories 
  defined by the continuation of the eigenvalues of the singlet anomalous dimension matrix~\eqref{eq:singlet-results} 
  (both continued from $s \in 2\mathbb{Z}_{>0}$).  
\end{itemize} 
The study of the analytic continuation of CFTs in spin (see~\cite{Caron-Huot:2017vep, 
Kravchuk:2018htv}) suggests that together with the Regge trajectories defined by the 
analytic continuation of the local operators, we should consider other types of trajectories: 
the shadow trajectories, defined by the shadow scaling dimensions, $\widetilde{\Delta}(s) = 
d - \Delta(s)$, and the nearly horizontal BFKL-like trajectories corresponding to specific 
values of spin, starting from a specific $s = J$, where $J$ is the value of the Regge 
intercept (see discussion in~\cite[Section 2.6]{Caron-Huot:2017vep}), and spanning an infinite series in the direction 
of negative spins.

\begin{figure}[h]
\centering
\includegraphics[width=0.6\columnwidth]{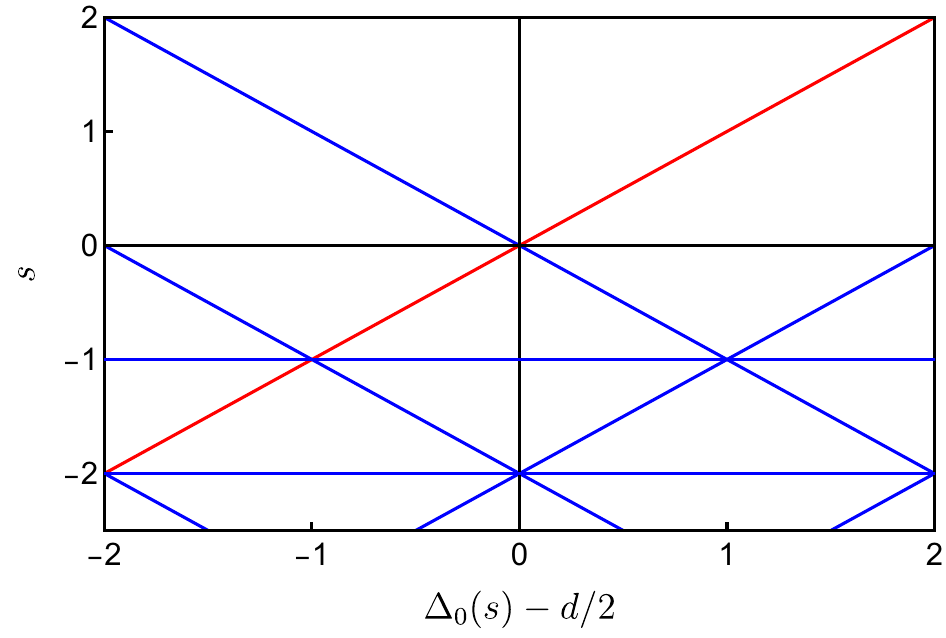}
\caption{Conformal Chew-Frautschi plot in the free theory ($\epsilon = 0$) in the 
chiral-even, $C$- and $P$-even symmetric-traceless sector. The red line represents the 
leading-twist trajectory, while the blue lines represent the other Regge trajectories, 
including trajectories for the operators of higher twist, shadow trajectories, and BFKL-like 
horizontal trajectories.}
\label{fig:chew-free}
\end{figure}

Let us start by considering the free conformal Regge trajectories corresponding to the 
family of leading-twist operators, $\Delta_{\tau = 2,0}(s) = 2 + s$. Here we provide the discussion for the generic case of an operator of twist 2 and return to the specific examples in Section~\ref{subsec:small-resum}. The schematic 
Chew-Frautschi plot including this trajectory is shown in Fig.~\ref{fig:chew-free}, where 
$\Delta_{\tau = 2,0}(s)$ is depicted in red. It is easy to see that the points where 
singularities are located, $s = 0, -1, -2, \ldots$, correspond to the intersection of 
several trajectories. In particular, at the point $s = 0$ the leading-twist trajectory 
intersects with its own shadow. Indeed, when $\epsilon = 0$
\begin{equation}
  \Delta_{\tau = 2,0}(0) = d - \Delta_{\tau = 2,0}(0) = 2. 
\end{equation}
At the point $s = -1$, the leading-twist trajectory intersects with the shadow trajectory 
for the operators of canonical twist four, $\tau = 4$, indeed, $\Delta_{\tau = 2,0}(-1) = 
d - \Delta_{\tau = 4, 0}(-1)$. Additionally, horizontal BFKL-like trajectories can 
intersect the leading-twist one at integer values of spin. Their construction is not possible 
in terms of local operators and, in principle, it is not obvious at which point they would 
contribute. We believe that starting from $s = -1$ one certainly has to take them into 
account and we return to the case $s = 0$ in what follows. Note also that we consider only the 
trajectories related to operators with the same quantum numbers as the leading-twist 
operators~\eqref{eq:nonsinglet-operator-def},~\eqref{eq:singlet-operator-def}, which explains why Fig.~\ref{fig:chew-free} does not contain twist-3 
trajectories.

Thus, the points where several free conformal Regge trajectories intersect can be interpreted 
as degeneracies in the spectrum of the free theory. It is well known~\cite{Korchemsky:2015cyx} 
that whenever we turn to the interacting theory, such degeneracies contribute as divergences 
in the naive perturbation theory (see e.g.~\cite{Lang:1993ct, Derkachov:1998js, 
Manashov:2026tcd} for more examples of such behavior). These divergent perturbative series, 
however, can be reorganized into a non-perturbative trajectory close to the point of 
intersection. Let us demonstrate this mechanism for the intersection of two trajectories, 
$\Delta_{\tau = 2}(s) = 2 - 2\epsilon + s + \gamma(s)$ and its shadow 
$\widetilde{\Delta}_{\tau = 2}(s) = 2 - s - \gamma(s)$. In the vicinity of the point where 
only two of them intersect, we expect the non-perturbative trajectory to satisfy the 
characteristic equation
\begin{equation}
\label{eq:char-equation-branching}
  \Delta^2 - \big(\Delta_{\tau = 2}(s) + \widetilde{\Delta}_{\tau = 2}(s)\big)\Delta + 
  \Delta_{\tau = 2}(s)\cdot \widetilde{\Delta}_{\tau = 2}(s) = 0.
\end{equation}    
When we are far from the point of intersection, $\Delta_{\tau = 2}(s) \ne 
\widetilde{\Delta}_{\tau = 2}(s)$, we have two non-degenerate solutions lying on different 
sheets of the Riemann surface. When we are close to the intersection, we have regular 
behavior but hit the branch points of the solution. If the 
equation~\eqref{eq:char-equation-branching} is correct, its coefficients, $\Delta_{\tau = 
2}(s) + \widetilde{\Delta}_{\tau = 2}(s)$ and $\Delta_{\tau = 2}(s)\cdot 
\widetilde{\Delta}_{\tau = 2}(s)$, should remain regular at the intersection point. For the 
sum, this statement is trivial, as $\Delta_{\tau = 2}(s) + \widetilde{\Delta}_{\tau = 
2}(s) = d$, while for the product it gives a non-trivial constraint.

\subsection{Resummation of the $s \to 0$ singularity \label{subsec:small-resum}}

In the previous section we introduced the equation~\eqref{eq:char-equation-branching}, 
which describes the square-root type of branching, and this section is devoted to using it 
to resum the $s \to 0$ singularities in the critical anomalous dimensions. The 
whole procedure relies on the fact that while anomalous dimensions contain poles at 
$s \to 0$, the product $\Delta(s)(d - \Delta(s))$ should remain regular. For convenience, 
we also drop the a priori regular part and introduce the following function
\begin{align}
  \label{eq:delm-def}
\delta m^2(s) = 2\gamma(s)\left(s - \epsilon + \dfrac{1}{2}\gamma(s)\right), && 
\Delta(s)\big(d - \Delta(s)\big) = (2- s)(2 + s - 2\epsilon) - \delta m^2(s).
\end{align} 
The origin of this notation can be found in~\cite{Manashov:2025kgf}. The resummed anomalous 
dimension in the vicinity of the point $s = 0$ is then defined as
\begin{equation}
\label{eq:resummed-anomalous-dimension}
\gamma(s)\big|_{s \to 0} = \epsilon - s + \sqrt{\big(\epsilon - s\big)^2 + \delta m^2(s)},
\end{equation}
where the square root is understood as a multivalued function defined on the appropriate 
two-sheeted Riemann surface.

\begin{figure}[h]
\centering
\begin{subfigure}[b]{0.45\columnwidth}
\centering
\includegraphics[height=0.65\columnwidth]{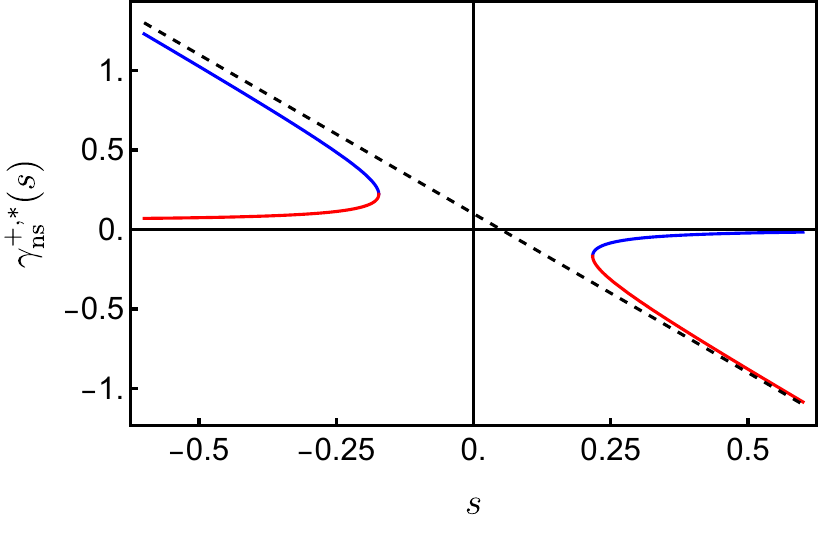}
\vspace{0.15cm}
\caption{}
\label{fig:ns-even-resum}
\end{subfigure}
\hfill
\begin{subfigure}[b]{0.5\columnwidth}
\centering
\includegraphics[height=0.8\columnwidth]{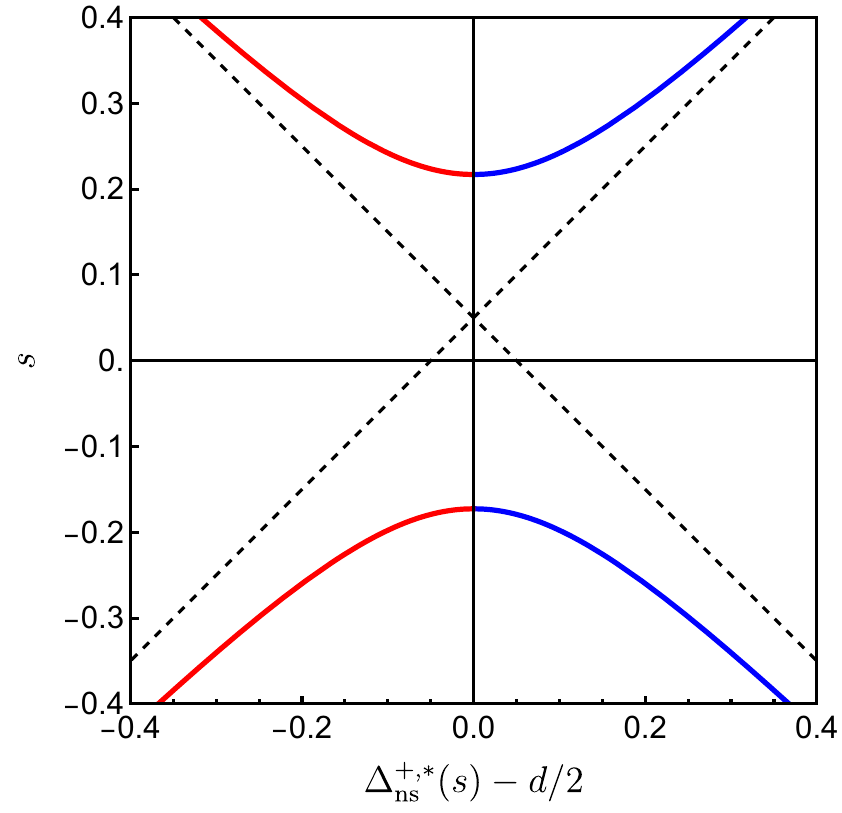}
\caption{}
\label{fig:ns-even-chew}
\end{subfigure}
\caption{(a) Resummation of the $s \to 0$ singularity for the non-singlet anomalous 
dimension continued from even values of spin, $\gamma^{+, *}_{\text{ns}}(s)$. 
(b) Corresponding conformal Chew-Frautschi plot. In both panels, red and blue curves 
represent the two branches of a square root, while the black dashed line corresponds to 
the free theory. The plot is made for $N = 1$, $\epsilon = 0.05$.}
\label{fig:ns-even}
\end{figure}

Let us check whether the resummation procedure works with the obtained results. First, 
consider the non-singlet anomalous dimension continued from even values of spin, 
$\gamma^{+, *}_{\text{ns}}(s)$. Constructing the corresponding function 
$\delta m^{2}_{\text{ns}, +}(s) = 2\gamma^{+, *}_{\text{ns}}(s)\big(s - \epsilon + 
\gamma^{+, *}_{\text{ns}}(s)/2\big)$, we find
\begin{align}
\label{eq:ns-even-delm}
\delta m^2_{\text{ns}, +}(0) &= -4u_1\epsilon + \Big((19 + 12N)u_1^2 - 4u_2\Big)\epsilon^2 
+ \bigg(-4u_3 + 2u_1u_2 + 18u_2 + \dfrac{17}{2}\lambda_1^2u_1 - 36\lambda_1u_1^2 \nonumber 
\\ &\quad +  3u_1 - \left(8\zeta_2 + \dfrac{235}{2}\right)u_1^2 + \left(\dfrac{44}{5}
\zeta_2^2 -60\zeta_3 + 20\zeta_2 + \dfrac{1143}{4}\right)u_1^3\bigg)\epsilon^3 + 
O(\epsilon^4),
\end{align}
which indeed makes the behavior of~\eqref{eq:resummed-anomalous-dimension} regular around
$s = 0$. Using the result~\eqref{eq:ns-even-delm} together with the resummation 
procedure~\eqref{eq:resummed-anomalous-dimension}, we plot the resulting anomalous 
dimension, $\gamma^{+, *}_{\text{ns}}(s)$, together with the corresponding 
Chew-Frautschi plot in Fig.~\ref{fig:ns-even}. We see that in the interacting theory, 
both the shadow and leading-twist trajectories resolve into smooth curves.

Remarkably, an attempt to apply the same resummation 
procedure~\eqref{eq:resummed-anomalous-dimension} to the non-singlet anomalous dimensions 
continued from odd values of spin fails. For the combination~\eqref{eq:delm-def} we get
\begin{align}
\label{eq:deltaMminus-sing}
\delta m_{\text{ns}, -}^2(s) &= 8u_1^2\dfrac{\epsilon^2}{s} + O(s^0,\epsilon^3),
\end{align}
which is still singular at $s = 0$. The natural possible explanation for this effect, which we can make in light of Section~\ref{subsec:regge}, is that there is another 
conformal Regge trajectory intersecting with $\Delta^{-, *}_{\text{ns}}(s)$ and 
$\widetilde{\Delta}^{-, *}_{\text{ns}}(s)$ at the point $s = 0$. The natural candidate 
for this is the horizontal trajectory at $s = 0$. Operators~\eqref{eq:nonsinglet-operator-def} continued from odd and even values of $s$ have different quantum numbers, e.g. $C$-parity. If our assumption about the existence of a trajectory at $s = 0$ is correct, 
the corresponding non-local operator should carry the same $C$-parity as the 
operator~\eqref{eq:nonsinglet-operator-def} with $s \in 2\mathbb{Z}_{>0} + 1$. We give 
additional details in Appendix~\ref{app:s0-bfkl}.

\begin{figure}[h]
\centering
\begin{subfigure}[b]{0.45\columnwidth}
\centering
\includegraphics[width=0.9\columnwidth]{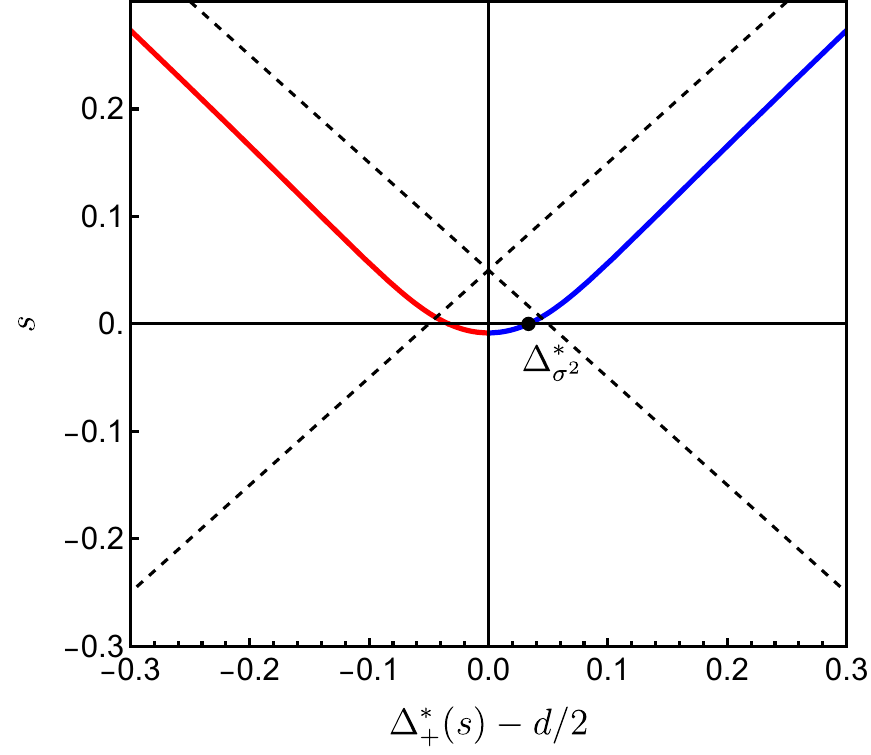}
\caption{}
\label{fig:s-positive-chew}
\end{subfigure}
\hfill
\begin{subfigure}[b]{0.45\columnwidth}
\centering
\includegraphics[width=0.9\columnwidth]{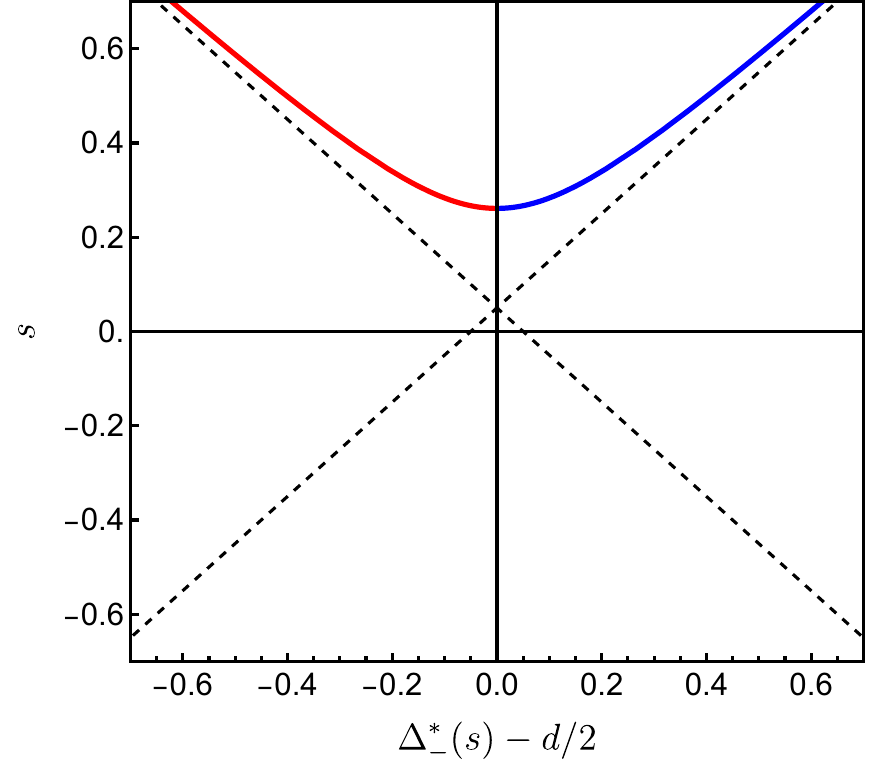}
\caption{}
\label{fig:s-negative-chew}
\end{subfigure}
\caption{Conformal Chew-Frautschi plot for the flavor-singlet trajectories around $s = 0$. 
Panels (a) and (b) correspond to the two different eigenvalues of the anomalous dimension 
matrix~\eqref{eq:eigenvalues-definition}. In both panels, blue and red curves represent the two branches of a 
square root. In panel (a), the position of the local operator $\sigma^2(x)$ is indicated 
by a black dot. The plot is made for $N = 1$, $\epsilon = 0.05$.}
\label{fig:s-chew}
\end{figure}

We also apply the resummation procedure to the singlet anomalous dimensions. Since the 
corresponding Regge trajectories are built from the eigenvalues of the anomalous dimension 
matrix, we construct the combinations analogous to~\eqref{eq:delm-def} and check that they 
are indeed regular at $s = 0$. Using the results~\eqref{eq:singlet-results} we find
\begin{align}
\label{eq:delm-positive}
\delta m^2_{+}(0) &= -2\big(\lambda_1^2 - 6u_1 + 18u_1^2\big)\epsilon^2 + 
\Big(24N(1 + 2u_1)u_2 - 4\lambda_2\lambda_1 + 32\lambda_1^2u_1 - 192\lambda_1u_1^2 
\nonumber \\ &\quad + 208\lambda_1u_1 - 48\lambda_1 + (8N -13)\lambda_1^2 + 
2(6N + 5)\lambda_1^3 - 196Nu_1^2 + 180Nu_1^3\Big)\epsilon^3 + O(\epsilon^4), \\
\label{eq:delm-negative}
\delta m^2_{-}(0) &= -4u_1\epsilon + \Big(-4u_2 + 6u_1 + u_1^2\Big)\epsilon^2 + 
\bigg(-4u_3 + (18 - 16N + 2u_1)u_2 - \dfrac{823}{12}\lambda_1^2u_1\nonumber \\ &\quad  
+ 348\lambda_1u_1^2 -272\lambda_1u_1 - 12N\lambda_1^3  + \left(\dfrac{70}{3} - 
8N\right)\lambda_1^2 + 48\lambda_1 + 51 u_1 - \left(8\zeta_2 + \dfrac{775}{2}\right)u_1^2 
\nonumber \\ &\quad + \left(\dfrac{44}{5}\zeta_2^2 - 60\zeta_3 + 20\zeta_2 + 
\dfrac{2655}{4}\right)u_1^3\bigg)\epsilon^3 + O(\epsilon^4).
\end{align}
Note that while the leading term in $\delta m^2_{-}(0)$ coincides with the first term of 
$\delta m^2_{\text{ns}, +}(0)$, the $O(\epsilon)$ term of $\delta m^2_{+}(0)$ is absent. 
This is a manifestation of an important structure. In the model~\eqref{eq:gny-action-def} 
there is an operator $\sigma^2(x)$, whose quantum numbers coincide with those of the 
flavor-singlet family~\eqref{eq:singlet-operator-def}, but which is excluded from this 
family because the quark-singlet operator $\mathcal{O}^{q}_{s = 0}(x)$ is not 
well-defined. However, we should expect to find this local operator as a point on the 
conformal Regge trajectory defined by the eigenvalues of the anomalous dimension matrix, 
and this is indeed what happens after resummation. Using 
formula~\eqref{eq:resummed-anomalous-dimension} and result~\eqref{eq:delm-positive}, we 
reproduce the critical anomalous dimension of the operator $\sigma^2(x)$ up to $O(\epsilon^2)$ 
(see~\eqref{eq:gamma-squared})
\begin{equation}
  \label{eq:positive-limit}
\gamma^*_{+}(s)\big|_{s \to 0} = \gamma^{*}_{\sigma^2} = \big(\lambda_1 + 
4Nu_1\big)\epsilon + \bigg(\lambda_2 + 4Nu_2 - 4N\lambda_1u_1 -\dfrac{5}{6}\lambda_1^2 - 
2Nu_1^2\bigg)\epsilon^2 + O(\epsilon^3).
\end{equation}
Using the results~\eqref{eq:delm-positive} and~\eqref{eq:delm-negative}, we plot the 
resummed conformal Regge trajectories corresponding to the flavor singlet sector in 
Fig.~\ref{fig:s-chew}. 

\begin{figure}[h]
\centering
\begin{subfigure}[b]{0.45\columnwidth}
\centering
\includegraphics[width=0.9\columnwidth]{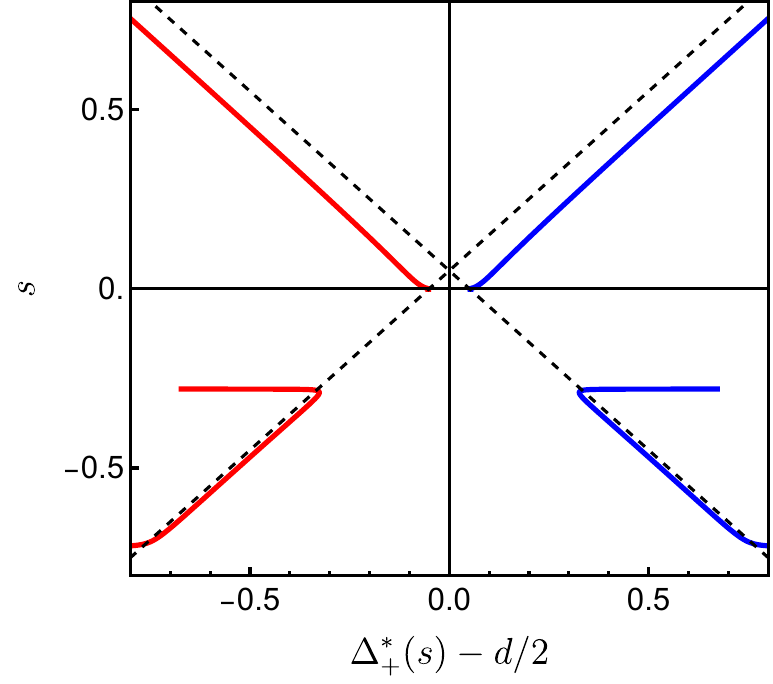}
\caption{}
\label{fig:branching25}
\end{subfigure}
\hfill
\begin{subfigure}[b]{0.45\columnwidth}
\centering
\includegraphics[width=0.9\columnwidth]{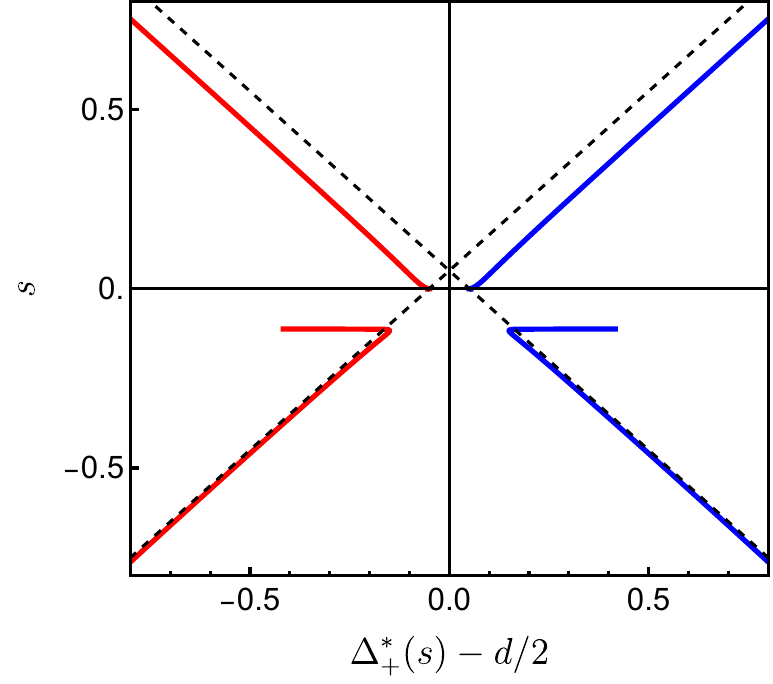}
\caption{}
\label{fig:branching50}
\end{subfigure}
\caption{Different positions of the mixing between $\Delta^{*}_{+}(s)$ and 
$\Delta^{*}_{-}(s)$ depicted on the conformal Chew-Frautschi plot for $\Delta^*_{+}(s)$. 
Panel (a) is made for $N = 25$ and panel (b) is made for $N = 50$. Both plots are made 
for $\epsilon = 0.05$.}
\label{fig:branching}
\end{figure}

Both plots show only the upper part of the trajectory. The reason is the presence of 
another branching due to the existing mixing between the trajectories $\Delta^{*}_{+}(s)$ 
and $\Delta^{*}_{-}(s)$. This branching manifests as square roots in the definition of 
the eigenvalues. The situation becomes clearer if one analyzes the trajectory for larger 
values of $N$, when the branching point moves towards $s = 0$, see 
Fig.~\ref{fig:branching}. In principle, it should be possible to construct a regular plot 
over the whole region $-1 < s < 1$ by unifying four trajectories: $\Delta^{*}_{+}(s)$, 
$\Delta^{*}_{-}(s)$, and both of their shadows, but we do not pursue this direction here 
and limit ourselves to the resummation of the singularity at $s = 0$.

At the end of this section let us mention the connection of the applied resummation 
procedure to a technique originating in the QCD context --- the double logarithmic 
equation (DLE). The original statement relies on the double-logarithmic asymptotics of the 
forward Compton scattering~\cite{Kirschner:1983di}, which at one loop yields the regular 
behavior of exactly the same combination of anomalous dimensions~\eqref{eq:delm-def}, 
see~\cite{Ermolaev:1995fx, Blumlein:1995jp} for details. The original one-loop DLE was 
later empirically generalized to $\mathcal{N} = 4$ SYM~\cite{Velizhanin:2011pb} and then, 
by adding the $\beta$-function contribution, to QCD~\cite{Velizhanin:2014dia}. The 
analysis of the conformal Chew-Frautschi plot (see~\cite{Chang:2025zib} for the one-loop 
analysis in QCD) places these generalizations on a more rigorous footing. Of particular 
interest is the case of odd $s$, analog of~\eqref{eq:deltaMminus-sing}. The double-logarithmic 
asymptotics in this case yields a considerably more involved regular combination,  
see~\cite[Eq. (13)]{Blumlein:1995jp}. It is interesting whether the arguments of 
Appendix~\ref{app:s0-bfkl} can be extended to a similar prediction for the case of odd 
spins.
 
\subsection{Regge intercepts and higher orders\label{subsec:small-use}}

In this section we argue that the possibility to effectively use the resummation procedure 
based on~\eqref{eq:char-equation-branching} can be seen not only as an exercise with the 
obtained results but also as having a certain practical implementation. We start here following 
the procedure described in~\cite[Section 2.5]{Caron-Huot:2017vep}. The correct behavior 
of the resummed spectrum of the analytically continued CFT around $\Delta(s) = d/2$ allows 
one to extract the values of the Regge intercepts, which can be determined as the largest 
values of spin, $s = J$, among all the trajectories at $\Delta(s) = d/2$. This value, $J$, 
controls the Regge limit of the correlators in the critical CFT and therefore is of direct 
interest (see~\cite{Costa:2012cb} for details).

From the computational side, the Regge intercept is simply defined as the position of the 
larger branching point of the square root~\eqref{eq:resummed-anomalous-dimension}. Let us 
start with the determination of the $\epsilon$-expansion of the intercept for the trajectory 
defined by $\gamma^{*}_{+}(s)$. Building an expansion of the form
\begin{equation}
\label{eq:regge-int-plus}
  J^{+}(\epsilon) = J_1^{+}\epsilon + J_2^{+}\epsilon^2 + J_3^{+}\epsilon^3 + O(\epsilon^4),
\end{equation} 
and expanding $\delta m^2_{+}(s)$ in a perturbative series, we obtain
\begin{subequations}
\begin{align}
  J_1^{+} &= \dfrac{1}{3 + 2N}\left(3 - 10N + \dfrac{2}{3}\varkappa\right), \\ 
  J_2^{+} &= \dfrac{1}{54\varkappa\delta (3 + 2N)^3}\bigg(-39168N^6 - 1288512N^5 + 163456N^4 + 3618480N^3 - 208224N^2 \nonumber \\ &\quad - 177768N - 3564 + \delta\Big(1543680N^5 + 1124640N^4 - 1597832N^3 + 177228N^2 + 39852N \nonumber \\ &\quad - 1188\Big)+ \varkappa\Big(768N^5 + 25472N^4 + 2640N^3 - 109512N^2 - 24084N - 1134 + \delta\big(-104064N^4 \nonumber \\ &\quad - 112672N^3 + 63672N^2 + 4302N - 378\big)\Big)\bigg), 
\end{align}
\end{subequations}
where we use the notation 
\begin{align}
  \varkappa = \sqrt{220N^2 - 156N + 9 + \delta(3 - 2N)}, && \delta  = \sqrt{4N^2+ 132N + 9}.
\end{align}
The available three-loop results allow one to extract the position of this Regge intercept 
up to $O(\epsilon^2)$. It is, however, possible to go further if we use the anomalous 
dimension of the operator $\sigma^2(x)$~\eqref{eq:gamma-squared} in view of the 
connection~\eqref{eq:positive-limit}. This allows one to determine $\delta m^2_{+}(0)$ at 
order $O(\epsilon^4)$ and obtain $J_3^{+}$. The result is quite lengthy and we provide it 
in an implicit form
\begin{equation}
\label{eq:regge-int-plus-3}
  J_{3}^{+} = \dfrac{1}{J_1^{+} - \frac{4\varkappa}{3(3 + 2N)}}\Big(m_{40} + 
  J^{+}_{1}m_{31} + \left(J_{1}^{+}\right)^2m_{22} + J_2^+ m_{21} + 
  \left(J_1^{+}\right)^3 m_{13} + 2J_2^+ J_1^+ m_{12} + \left(J_2^{+}\right)^2\Big),
\end{equation}
where
\begin{subequations}
\begin{align}
  m_{12} &= 8N\dfrac{1 - 12N}{3 + 2N}, \\
  m_{13} &= 4N\dfrac{240N^2 - 56N + 3}{3 + 2N}, \\
  m_{21} &= \dfrac{1}{27(3 + 2N)^3}\Big(384N^4 - 3104N^3 - 6648N^2 - 8190N + 378 \nonumber \\  &\quad+ \delta\big(-192N^3 - 32N^2 + 828N + 126\big)\Big), \\ 
  m_{22} &= \dfrac{4}{27(3 + 2N)^3}\Big(-1536N^5 + 2432N^4 + 8384N^3 + 9633N^2 + 828N - 81 \nonumber \\  &\quad + \zeta_2\big(864N^3 + 1728N^2 + 648N\big) + \delta\big(768N^4 - 64N^3 - 2110N^2 + 123N - 27\big) \Big), \\
  m_{31} &= \dfrac{1}{243\delta(3 + 2N)^5}\Big\{-12288N^8 - 405504N^7 + 327552N^6 - 1793824N^5 - 13745952N^4 \nonumber \\ &\quad - 20106216N^3 - 26826012N^2 - 1035180N + 119799 + \big(4608N^6 + 214272N^5 + 2198016N^4 \nonumber \\ &\quad + 4634496N^3 + 1796256N^2 - 1434672N - 104976\big)\zeta_2 + \delta\Big(6144N^7 + 101376N^6 + 206016N^5  \nonumber \\ &\quad + 157040N^4 + 1682040N^3 + 2301858N^2 + 499338N + 39933 + \big(186624N^4 + 606528N^3 \nonumber \\ &\quad + 839808N^2 + 524880N\big)\zeta_3 + \big(-2304N^5 + 55296N^4 + 196992N^3 + 171072N^2 - 11664N  \nonumber \\ &\quad- 34992\big)\zeta_2\Big)\Big\}, \\
  m_{40} &= \dfrac{1}{729\delta^3(3 + 2N)^6}\Big\{-92160N^9 - 6231808N^8 - 108825728N^7 + 587524800N^6 + 3292504416N^5 \nonumber \\ &\quad - 3717922464N^4 - 8582440536N^3 - 2255120676N^2 - 219550014N - 470205 + \big(-147456N^9 \nonumber \\ &\quad - 7852032N^8 - 7133184N^7 + 3385857024N^6 + 12747538944N^5 + 17096811264N^4 \nonumber \\ &\quad + 8108346240N^3 - 376653888N^2 + 24564384N + 5668704\big)\zeta_3 + \delta\Big(46080N^8 + 2635520N^7 \nonumber \\ &\quad + 32893984N^6 - 123577920N^5 + 184919868N^4 - 32777244N^3 + 682169769N^2 + 44528778N \nonumber \\ &\quad - 156735 + \big(73728N^8 + 2709504N^7 + 2405376N^6 - 252315648N^5 - 893182464N^4 \nonumber \\ &\quad - 982855296N^3 - 301910976N^2 + 11337408N + 1889568\big)\zeta_3\Big)\Big\}.
\end{align}
\end{subequations}

For the two other trajectories, defined by $\gamma_{\text{ns}}^{+, *}(s)$ and 
$\gamma_{-}^{*}(s)$, the $\epsilon$-expansion of the Regge intercept proceeds in powers of 
$\sqrt{\epsilon}$ and we build an expansion
\begin{align}
  J^{\text{ns}}(\epsilon) = J^{\text{ns}}_1\epsilon^{1/2} + J^{\text{ns}}_2\epsilon + 
  J^{\text{ns}}_3\epsilon^{3/2} + O(\epsilon^2), && J^{-}(\epsilon) = J^{-}_1\epsilon^{1/2} 
  + J^{-}_2\epsilon + J^{-}_3\epsilon^{3/2} + O(\epsilon^2).
\end{align}
Using the results~\eqref{eq:ns-results} and~\eqref{eq:singlet-results} we obtain
\begin{subequations}
\begin{align}
  J^{\text{ns}}_1 &= \dfrac{2}{\sqrt{3 + 2N}}, \\
  J^{\text{ns}}_2 &= \dfrac{2N}{3 + 2N}, \\ 
  J^{\text{ns}}_3 &= \dfrac{1}{108(3 + 2N)^{5/2}}\Big(-1304N^2 - 1560N - 531 + \delta\left(4N + 66\right)\Big), \\
  J^{\text{ns}}_4 &= \dfrac{1}{18(3 + 2N)^3}\Big(544N^2 + 375N - 99 + \big(216N + 324\big)\zeta_2 + \delta\left(-2N - 33\right)\Big), \\ 
  J^{\text{ns}}_5 &= \dfrac{1}{233280(3 + 2N)^{9/2}\delta}\Big(-298240N^5 - 16579200N^4 - 187907040N^3 + 51660720N^2 \nonumber \\ &\quad  - 115807320N + 8033580 + \delta\Big(-784000N^4 - 51091200N^3 - 87888240N^2 \nonumber \\ &\quad  - 26805600N + 10693215 + \big(9331200N^3 + 17729280N^2 - 9797760N - 23094720\big)\zeta_2 \nonumber \\ &\quad  + \big(-2052864N^2 - 6158592N - 4618944\big)\zeta_2^2 \nonumber \\ &\quad  + \big(-22394880N^2 - 55987200N - 33592320\big)\zeta_3\Big)\Big), \\
  J^{\text{ns}}_6 &= \dfrac{1}{6480(3 + 2N)^5\delta}\Big(36480N^5 + 2090240N^4 + 26720160N^3 + 6702480N^2 + 10357200N \nonumber \\ &\quad  - 669060 + \big(-253440N^4 - 8778240N^3 - 14307840N^2 - 2643840N - 116640\big)\zeta_2 \nonumber \\ &\quad + \delta\Big(-355200N^4 + 3616880N^3 + 6250680N^2 + 329940N - 1622700 + \big(-2102400N^3 \nonumber \\ &\quad - 4764960N^2 - 1723680N + 1040040\big)\zeta_2 + \big(67392N^2 + 202176N + 151632\big)\zeta_2^2 \nonumber \\ &\quad + \big(-207360N^3 + 1503360N^2 + 4976640N + 3382560\big)\zeta_3 \nonumber \\ &\quad + \big(-129600N^2 - 388800N - 291600\big)\zeta_5\Big)\Big),
\end{align}
\end{subequations}
for the non-singlet trajectory and
\begin{subequations}
\label{eq:regge-int-minus}
\begin{align}
  J^{-}_1 &= \dfrac{2}{\sqrt{3 + 2N}}, \\
  J^{-}_2 &= \dfrac{10N}{3 + 2N}, \\ 
  J^{-}_3 &= \dfrac{1}{108(3 + 2N)^{5/2}}\Big(-15552N^3 - 22904N^2 + 1032N - 531 + \delta\left(4N + 66\right)\Big), \\
  J^{-}_4 &= \dfrac{1}{54(3 + 2N)^3}\Big(104064N^4 + 112576N^3 - 58824N^2 + 3861N - 297 + \big(648N + 972\big)\zeta_2 \nonumber \\ &\quad  + \delta\left(-192N^3 + 16N^2 + 690N - 99\right)\Big),\\ 
  J^{-}_5 &= \dfrac{1}{233280(3 + 2N)^{9/2}\delta}\Big(179159040N^7 + 6161909760N^6 + 8077617920N^5 - 18845696640N^4 \nonumber \\ &\quad - 27100954080N^3 - 467620560N^2 - 31826520N + 8033580 + \delta\Big(-14646251520N^6 \nonumber \\ &\quad - 35337047040N^5 - 8412964480N^4 + 16092040320N^3 - 2038808880N^2 + 7486560N \nonumber \\ &\quad + 10693215 + \big(59719680N^4 + 240744960N^3 + 308862720N^2 + 107775360N - 23094720\big)\zeta_2 \nonumber \\ &\quad + \big(-2052864N^2 - 6158592N - 4618944\big)\zeta_2^2 \nonumber \\ &\quad + \big(-22394880N^2 - 55987200N - 33592320\big)\zeta_3\Big)\Big)\\
  J^{-}_6 &= \dfrac{1}{19440(3 + 2N)^5}\Big(-363233280N^8 - 12436439040N^7 - 14493895680N^6 + 38169330560N^5 \nonumber \\ &\quad + 44192017920N^4 - 8325920160N^3 - 514535760N^2 + 75615120N - 2007180 \nonumber \\ &\quad + \big(-760320N^4 - 26334720N^3 - 42923520N^2 - 7931520N - 349920\big)\zeta_2 \nonumber \\ &\quad + \delta\Big(22128599040N^7 + 49718292480N^6 + 3657753600N^5 - 27161340160N^4 + 6476169840N^3\nonumber \\ &\quad - 467956440N^2 + 46834740N - 4868100 + \big(-49766400N^5 - 171694080N^4 - 193553280N^3 \nonumber \\ &\quad - 88944480N^2 - 23366880N + 3120120\big)\zeta_2 + \big(-1368576N^3 - 3903552N^2 \nonumber \\ &\quad - 2472768N + 454896\big)\zeta_2^2 + \big(2488320N^4 + 2488320N^3 + 4510080N^2 + 16329600N \nonumber \\ &\quad + 10147680\big)\zeta_3 + \big(-388800N^2 - 1166400N - 874800\big)\zeta_5\Big)\Big),
\end{align}
\end{subequations}
for the singlet one. 

After building these expansions, we see that the Regge intercepts $J^{+}(\epsilon)$ and 
$J^{-}(\epsilon)$ have different scaling ($\sim\epsilon$ vs $\sim\sqrt{\epsilon}$) near 
$d = 4$. This allows us to associate $J^{-}(\epsilon)$, Eq.~\eqref{eq:regge-int-minus}, 
with the Pomeron of the critical CFT. This, however, can change if one builds a suitable 
resummation to the point $d = 3$, so both values~\eqref{eq:regge-int-plus} 
and~\eqref{eq:regge-int-minus} are worth mentioning. It would be highly desirable to 
cross-check the Pomeron intercept against results obtained by other methods (e.g. by 
conformal bootstrap techniques as in the Ising case~\cite{Caron-Huot:2020ouj}), but such 
results are not currently known to the authors. The Regge intercept of the non-singlet 
trajectory, $J^{\text{ns}}(\epsilon)$, controls the non-singlet channel in the Regge limit.

Another application of the resummation procedure performed in Section~\ref{subsec:small-resum} 
is related to predictions for higher-loop orders. The regularity of the 
combination~\eqref{eq:delm-def} requires that the singular expansions of the critical 
anomalous dimensions at higher-loop orders depend on the lower-loop data. For example, at 
$O(\epsilon^4)$ we can require that the combination
\begin{equation}
\label{eq:gamma-4-sing-def}
  (\delta m^2)^{(4)}(s) = 2s\gamma^{(4)}(s) + 2\gamma^{(3)}(s)\big(\gamma^{(1)}(s) - 1\big) 
  + \big(\gamma^{(2)}(s)\big)^2 
\end{equation}
is regular at $s = 0$. This fixes the asymptotic expansion of $\gamma^{(4)}(s)$ around 
$s = 0$ up to $O(1/s^2)$ terms. If the trajectory accommodates a certain local operator 
(see~\eqref{eq:positive-limit} for an example) and its anomalous dimension is known at 
order $O(\epsilon^3)$, one can additionally fix $\left(\delta m^2\right)^{(4)}(0)$ using 
this information and obtain a prediction for $\gamma^{(4)}(s)$ at order $O(1/s)$. This is 
similar to the idea used in the calculation~\eqref{eq:regge-int-plus-3}.

To give an example, we provide here a few leading terms of the critical anomalous 
dimension $\big(\gamma^{+}\big)^{(4)}(s)$ at order $O(\epsilon^4)$:
\begin{align}
\label{eq:gamma-4-prediction}
  \big(\gamma^{+}\big)^{(4)}(s) &= \bigg(-492832N^4 + 654720N^3 - 273888N^2 + 38448N - 
  1782 \nonumber \\ &\quad  + \delta\left(5360N^3 - 11448N^2 + 5508N - 594\right)\bigg)
  \dfrac{1}{81(3 + 2N)^4s^3} + \bigg(-1761792N^7 \nonumber \\ &\quad  - 57385600N^6 + 
  26102592N^5 + 162387216N^4 - 60862536N^3 - 9386604N^2  \nonumber \\ &\quad  + 1507086N 
  + 52488 + \delta\Big(57801216N^6 + 23800256N^5 - 73939344N^4 + 24074280N^3  \nonumber 
  \\ &\quad + 621972N^2 - 439506N + 17496\Big)\bigg)\dfrac{1}{243(3 + 2N)^5\delta s^2} 
  + O(1/s).
\end{align}
The expression at order $O(1/s)$ is quite lengthy and we do not provide its exact form 
here, but it can be obtained using~\eqref{eq:gamma-4-sing-def}. We also do not 
provide the leading asymptotic expansions for the critical anomalous dimensions 
$\gamma^*_{+}(s)$ and $\gamma^*_{-}(s)$, only mentioning that they can be predicted up to 
order $O(1/s^2)$.  

The knowledge of the singular contributions to the higher-order anomalous dimensions can 
be seen as a certain improvement to the LLL-based reconstruction procedure described in 
Section~\ref{subsec:method}. Indeed, each coefficient in~\eqref{eq:gamma-4-prediction} 
potentially gives an additional equation to the set of Diophantine equations without the 
calculation of an additional data point. 
Note that the same comment about the 
limit $\lambda \to 0$ as in Section~\ref{subsec:reciprocity} applies here. While the 
prediction of the singular behavior for the critical anomalous dimensions has limited 
applications for calculations at generic $(u, \lambda)$, working in the limit $\lambda \to 
0$ allows one to directly reproduce the singular contribution to $u^4$ at four loops, 
which is the most technically demanding piece.

\section{Summary\label{sec:sum}}

In this paper, we presented the calculation of the three-loop anomalous dimensions of the 
leading-twist operators in the Gross-Neveu-Yukawa model. The calculation was done with the 
help of the technique developed in~\cite{Chakraborty:2026lna}. We generalized the model 
to an arbitrary number of scalar and fermionic degrees of freedom, $N_s$ and $N_f$, and 
used an emergent supersymmetry at specific values of $N_s$ and $N_f$ to find relations 
among different Feynman integrals in the flavor-singlet sector. This optimization allowed 
us to speed up the calculations by $\sim 25\%$, which at three-loop level led to saving 
days of computation. The use of the generalized Lagrangian also allowed us to obtain 
results for the leading-twist operators in the singlet sector of Nambu-Jona-Lasinio-Yukawa and Wess-Zumino 
models at negligible additional cost.

Of particular interest is the relation between emergent supersymmetry and analytic 
continuation in the number of dimensions. As the supersymmetric point, $N = 1/4$, 
corresponds to the 3D $\mathcal{N} = 1$ Wess-Zumino model, we verified 
explicitly that in order to observe the emergence of SUSY one has to add contributions 
from traces of an odd number of Dirac matrices, a distinguishing feature of 
3D theories. In the leading-twist anomalous dimensions this effect first 
appears at three loops. These studies provide a possible playground for the implementation of 
emergent SUSY in gauge theories and, in particular, in QCD calculations. There the main 
obstacle to using the emergent SUSY framework is the presence of $\gamma_5$, which also 
stems from the problems of theories in non-integer dimensions, 
see~\cite{Chakraborty:2026wie} for details.

The results pass a set of comprehensive checks. Most of them, such as the agreement with the 
$1/N$ expansion, reciprocity, and the small-spin limit, are tied to the critical properties 
of the GNY model. In this view, the properties related to the analytical structure of 
anomalous dimensions as functions of spin can also be seen as subjects for a deeper 
understanding of analytic continuation in spin in fermionic CFTs. In this paper we mostly 
follow the studies initiated at two loops in~\cite{Manashov:2025kgf} and thoroughly study 
the behavior of the critical anomalous dimensions around the rightmost singularity, $s = 0$. 
We find that the expected structure of the conformal Chew-Frautschi plot works particularly well in predicting analytical properties of anomalous dimensions. It would be interesting to perform a deeper study of the same structures 
through the lens of non-local detector operators in the GNY model. The study of the 
singular behavior around $s = 0$ is interesting not only from a conceptual perspective, 
but also allows us to obtain quantitative results. In particular, we extract the values of 
the Regge intercepts for all types of leading-twist trajectories in the form of 
$\epsilon$ expansions and predict the higher-order singular behavior of the critical 
anomalous dimensions.

Thus, the obtained results are interesting from two perspectives. On the one hand, they 
represent a high-order non-trivial calculation valuable for the critical properties of the 
GNY model. On the other hand, the tools used and the properties of the results studied 
have good potential to be exploited in more technically demanding calculations. In 
particular, the presence of the flavor-singlet sector establishes the GNY model as a 
natural testing ground for the analogous sector in QCD, which is tempting to push 
towards four-loop precision. 

\section*{Acknowledgments} We are grateful to Alexander Manashov and Sven-Olaf Moch for numerous illuminating discussions and critical remarks. We are also grateful to Vitaly Velizhanin and Andreas Vogt for discussions concerning the implementation of the LLL algorithm for the analytic reconstruction of anomalous dimensions. This work was supported by the ERC Advanced Grant 101095857 {\it Conformal-EIC}. 
\addcontentsline{toc}{section}{Acknowledgments}
\appendix
\section*{Appendices}

\addcontentsline{toc}{section}{Appendices} 
\section{Horizontal trajectory at $s = 0$ \label{app:s0-bfkl}}

In this appendix we analyze the possible non-local operator corresponding to the horizontal 
line, $s = 0$, in the conformal Chew-Frautschi plot, Fig.~\ref{fig:chew-free}, for the 
flavor non-singlet sector. In particular, we are interested in the possible mixing of such 
operators with the operator~\eqref{eq:nonsinglet-operator-def} for odd and even values 
of $s$.

For convenience we switch here to the free four-dimensional theory in Minkowski space, 
which is sufficient for the analysis of operator properties, and use spinor formalism. Our 
notation is the following:
\begin{align}
  \label{eq:spinor-formalism-def}
  q = \begin{pmatrix}
    \psi_{\alpha} \\ 
    \bar{\chi}^{\dot{\alpha}}
  \end{pmatrix}, && \bar{q} = \begin{pmatrix}
    \chi^{\alpha} & \bar{\psi}_{\dot{\alpha}}
  \end{pmatrix}, && \gamma^{\mu} = \begin{pmatrix}
    0 & (\sigma^{\mu})_{\alpha\dot{\beta}} \\ 
    (\bar{\sigma}^{\mu})^{\dot{\alpha}\beta} & 0 
  \end{pmatrix}.
\end{align}
The light-cone vector, $n^{\mu}$, in spinor formalism factorizes into the product of two 
spinors, $n_{\alpha\dot{\alpha}} = \lambda_{\alpha}\bar{\lambda}_{\dot{\alpha}}$, and the 
Weyl spinors entering the Dirac fermion~\eqref{eq:spinor-formalism-def} enjoy the 
light-cone projection
\begin{align}
  \psi_{+} = \lambda^{\alpha}\psi_{\alpha}, && \chi_{+} = 
  \chi^{\alpha}\lambda_{\alpha}, && \bar{\psi}_{+} = \bar{\psi}_{\dot{\alpha}}\bar{\lambda}^{\dot{\alpha}}, && \bar{\chi}_{+} = 
  \bar{\lambda}_{\dot{\alpha}}\bar{\chi}^{\dot{\alpha}}.
\end{align}
We use the same notation for the light-cone projection of vectors, e.g. $x_{+} = 
n^{\mu}x_{\mu}$.

The non-singlet fermion operator~\eqref{eq:nonsinglet-operator-def} in spinor formalism 
then takes the form
\begin{equation}
  \label{eq:non-singlet-spinor}
  \mathcal{O}^{q,A}_{s} = \bar{\psi}_{+, i}\,\partial_{+}^{s - 1}\tau_{ij}^A\psi_{+,j} + 
  \chi_{+, i}\,\partial_+^{s- 1}\tau_{ij}^A \bar{\chi}_{+, j}.
\end{equation}
The main difference between the cases of even and odd $s$ in the 
definition~\eqref{eq:non-singlet-spinor} is the behavior under charge conjugation. Indeed, 
one easily finds~\footnote{Here we assume $(\tau^A)^T = \tau^A$. If index $A$ denotes an antisymmetric matrix, this gives additional overall sign for all operators, which does not change any conclusions.}
\begin{equation}
  \mathcal{C}\,\mathcal{O}^{q,A}_{s}\,\mathcal{C}^{-1} = (-1)^s\mathcal{O}^{q,A}_{s},
\end{equation}
where $\mathcal{C}$ is the charge conjugation matrix, and for convenience we list how the 
light-cone projections of the Weyl spinors transform
\begin{align}
\label{eq:c-parity-rule}
  C\text{-parity:} && \psi_{+} \mapsto -\chi_{+},&& \chi_{+} \mapsto - \psi_{+}, && \bar{\psi}_+ 
  \mapsto - \bar{\chi}_{+}, && \bar{\chi}_{+} \mapsto - \bar{\psi}_{+}.
\end{align}
Thus, the operator~\eqref{eq:non-singlet-spinor} is $C$-even for even $s$ and $C$-odd for 
odd $s$.


Now let us address the question of what kind of operators correspond to the horizontal 
trajectories on the conformal Chew-Frautschi plot and, in particular, to the line $s = 0$. 
Here we adopt the construction described in~\cite[Section 2.6]{Caron-Huot:2022eqs}. The 
key building block for this is the light-ray transform~\cite{Kravchuk:2018htv}, which 
applied to the primary operator 
$\mathcal{O}_{\Delta, \alpha_1\ldots\alpha_{j_L}\dot{\alpha}_1\ldots\dot{\alpha}_{j_R}}$ 
with scaling dimension $\Delta$ reads as 
\begin{equation}
  \mathbf{L}\big[\mathcal{O}_{\Delta}\big](x, n) = 
  \int_{-\infty}^{+\infty}d\alpha\,(-\alpha)^{-\Delta - s}\lambda^{\alpha_1}\ldots
  \lambda^{\alpha_{j_L}}\bar{\lambda}^{\dot{\alpha_1}}\ldots\bar{\lambda}^{\dot{\alpha}_{j_R}}
  \mathcal{O}_{\Delta, \alpha_1\ldots\alpha_{j_L}\dot{\alpha}_1\ldots\dot{\alpha}_{j_R}}
  \left(x - \dfrac{n}{\alpha}\right),
\end{equation}
where $s = (j_L + j_R)/2$. The light-ray transform yields a primary operator on the celestial 
sphere with spin and scaling dimension $(\Delta_L, s_L) = (1 - s, 1 - \Delta)$. This 
property makes it possible to associate the product of $m$ different operators with a local 
spin $S$:
\begin{align}
  :\mathbf{L}\big[\mathcal{O}_1\big]\ldots\mathbf{L}\big[\mathcal{O}_m\big]:, && S = s_1 
  + \ldots + s_m - m + 1, 
\end{align}
where $s_k$ is the spin of the operator $\mathcal{O}_k$. If the resulting spin is less 
than the value of the Regge intercept (see Section~\ref{subsec:regge}), $S < J$, in an interacting theory this 
product requires renormalization and contributes to the conformal Chew-Frautschi plot.

Here we consider the case $m = 2$, since integrating the product 
$:\mathbf{L}\big[\mathcal{O}_1\big]\mathbf{L}\big[\mathcal{O}_2\big]:$ with an appropriate 
kernel (Clebsch-Gordan coefficient) leads to non-local operators spanning the entire 
horizontal line at $S = s_1 + s_2 - 1$. The presence of half-integer spin operators in 
the fermionic model allows us to accommodate $S = 0$ with the simplest possible building 
blocks. This is an important structural difference from the bosonic theories, where the simplest operators are given by scalars. Following the argument in~\cite{Caron-Huot:2022eqs} that the light-ray transform 
of a single basic field must be zero, we start with operators of the type 
$\mathbf{L}\big[\sigma \psi_{+}\big]$.~\footnote{Here, the operator $\mathbf{L}\big[\bar{q}q\psi_{+}\big]$ is also possible, but as $\bar{q}q$ behaves as scalar it does not change the conclusions of the discussion. Remarkably, this construction then can be translated to QCD, where scalar field $\sigma$ is absent.} One then constructs the following minimal operator
\begin{align}
  \label{eq:detector-hP}
  \mathcal{H}^{A, +}_{J_L}(x, n) &= \int D^{2}n_1 D^{2}n_2\,K_{J_L}(n_1, n_2; n)\Big[:\mathbf{L}\big[\sigma \bar{\psi}_{+, i}\big](x, 
  n_1)\tau^A_{ij}\mathbf{L}\big[\psi_{+,j}\sigma\big](x,n_2): \nonumber \\&\quad + :\mathbf{L}\big[\sigma \chi_{+, i}\big](x, 
  n_1)\tau^A_{ij}\mathbf{L}\big[\bar{\chi}_{+,j}\sigma\big]
  (x, n_2):\Big].
\end{align}
Here $D^2n$ is the measure on the celestial sphere and $K_{J_L}(n_1, n_2; n)$ is the Clebsch-Gordan coefficient connecting representations $(1/2, 0)$, $(0, 1/2)$ and representation with the spin $J_L$. The explicit expressions for both are not relevant for the present discussion. This integration associates the operator~\eqref{eq:detector-hP} with the whole horizontal line at $s = 0$, rather than single point. 

Using the rules~\eqref{eq:c-parity-rule} one can easily verify that operator~\eqref{eq:detector-hP} is $C$-odd, and therefore the corresponding trajectory is allowed to mix with the 
leading-twist trajectory continued from odd values of $s$. This can serve as a possible 
explanation for the presence of the singularity in~\eqref{eq:deltaMminus-sing}. The 
$C$-even combination,
\begin{align}
    \mathcal{H}^{A, -}_{J_L}(x, n) &= \int D^{2}n_1 D^{2}n_2\,K_{J_L}(n_1, n_2; n)\Big[:\mathbf{L}\big[\sigma \bar{\psi}_{+, i}\big](x, 
  n_1)\tau^A_{ij}\mathbf{L}\big[\psi_{+,j}\sigma\big](x,n_2): \nonumber \\&\quad - :\mathbf{L}\big[\sigma \chi_{+, i}\big](x, 
  n_1)\tau^A_{ij}\mathbf{L}\big[\bar{\chi}_{+,j}\sigma\big]
  (x, n_2):\Big],
\end{align} 
is at the same time $P$-odd and would mix with the axial operator 
$\bar{q}_i\gamma_{+}\gamma_5\partial_{+}^{s-1}\tau_{ij}^{A}q_j$ continued from odd
values of $s$. Remarkably, this expected behavior is consistent with the relations between non-singlet axial and vector anomalous dimensions in
QCD~\cite{Mertig:1995ny, Vogelsang:1996im}. 

Thus, this consideration is intended to show that the attempt to accommodate the horizontal 
line $s = 0$ leads to a detector operator mixing with the 
operator~\eqref{eq:nonsinglet-operator-def} continued from odd values of $s$. This 
probably explains the presence of the unresolved singularity in~\eqref{eq:deltaMminus-sing}, 
but the actual check would consist of a renormalization of the 
operator~\eqref{eq:detector-hP}. This also by no means serves as a proof that such a 
trajectory does not exist at $s = 0$ for the operator continued from even values of $s$, 
as we have by far not exhausted all possibilities for detector construction. The actual 
calculations, however, suggest that such a mixing is largely suppressed, if it even exists, since taking into 
account only two trajectories, the leading twist and its shadow, resolves the 
singularities in~\eqref{eq:ns-even-delm} up to the comparatively high perturbative order 
$O(\epsilon^3)$.
\renewcommand{\theequation}{\Alph{section}.\arabic{equation}}
\renewcommand{\thetable}{\Alph{table}}

\section{Generalised Lagrangian results}\label{app:lgen}
Here we present the results of our calculation done in the generalised Lagrangian model, which was crucial for the computational optimisation via Emergent SUSY. The number of scalars and fermions here are denoted as $N_s,N_f$ respectively.  The 3D enhancement is denoted by the symbol $\eta$, with $\eta=1$ for $d=3,N_s=1$ and $0$ for $d=4$. Throughout, we abbreviate
\begin{equation}
C \;\equiv\; N_s^2-6N_s+4 ,
\label{eq:Cdef}
\end{equation}
which appears as an overall factor in every $\zeta_3$ term and throughout the
weight-3 sector, and is therefore kept outside the brackets below. See \cite{Chakraborty:2026lna} for the field anomalous dimensions for $\lgen$. From the below results, we can get results for singlet operators in multiple models by simply substituting for $N_s,N_f$, as was shown in Table~\ref{tab:models-text}.
\begin{align}
\gamma_{qq}^{(3)} = {}
&2\gamma_q^{(3)} \;+\;\dfrac{u\lambda^2\,N_s (N_s+2)}{s(s+1)}\biggl[ -\frac{2 s+1}{18 s(s+1)} +\frac{17}{72}\biggr] \notag\\
&+\dfrac{2u^2\lambda\,N_s (N_s+2)}{s(s+1)}\biggl[ \frac{4}{3}S_{1} +\frac{2 (2 s+1)}{3 s(s+1)} -3\biggr] \notag\\
&-\dfrac{u^3\,N_s}{s(s+1)}\Bigg[ -\bigg(\frac{4 C}{s(s+1)} +4 C\bigg)S_{3} -\bigg(\frac{8 C}{s(s+1)} +8 C\bigg)\bigg(S_{1,-2}-\tfrac12 S_{-3}\bigg) \notag\\
&\quad +\bigg(\frac{8 C}{s^2(s+1)^2} +\frac{4 \left(5 N_s^2-26 N_s+16\right)}{s(s+1)}\bigg)S_{-2} \notag\\
&\quad +\bigg(16 +8 N_f^2 -8 N_s -2 N_f N_s +N_s^2\bigg)S_1^2 \notag\\
&\quad -\bigg(\frac{2 (N_s-4) N_s}{s(s+1)} -32 -48 N_f -8 N_f^2 +20 N_s +14 N_f N_s -3 N_s^2\bigg)S_{2} \notag\\
&\quad -\bigg(\frac{4 (N_s-4) N_s (2 s+1)}{s^2(s+1)^2} \notag\\
&\quad \ \ -\frac{1}{s(s+1)}\bigg(16 N_f^2 s +32 N_f^2 -16 N_f N_s s -36 N_f N_s +48 N_f s \notag\\
&\quad  \phantom{+\frac{1}{s^2(s+1)^2}\bigg(\:} +96 N_f +6 N_s^2 s +14 N_s^2 -40 N_s s -60 N_s +64 s +48\bigg) \notag\\
&\quad \ \ +\frac{1}{2}\bigg(128 N_f^2-108 N_f N_s+288 N_f+41 N_s^2-248 N_s+288\bigg)\bigg)S_{1} \notag\\
&\quad +\bigg(\frac{12 C}{s(s+1)}\bigg)\zeta_3 +\frac{4 N_s^2}{s^4(s+1)^4} \notag\\
&\quad -\frac{1}{s^3(s+1)^3}\bigg(-48 N_f N_s s-44 N_f N_s+18 N_s^2 s+4 N_s^2-24 N_s\bigg) \notag\\
&\quad +\frac{1}{s^2(s+1)^2}\bigg(48 N_f^2 s +24 N_f^2 -164 N_f N_s s -56 N_f N_s +160 N_f s +128 N_f +50 N_s^2 s \notag\\
&\quad  \phantom{+\frac{1}{s^2(s+1)^2}\bigg(\:} +6 N_s^2 -144 N_s s -16 N_s +32\bigg) \notag\\
&\quad -\frac{1}{2 s(s+1)}\bigg(224 N_f^2 s +160 N_f^2 -244 N_f N_s s -212 N_f N_s +736 N_f s +432 N_f +65 N_s^2 s \notag\\
&\quad  \phantom{+\frac{1}{s^2(s+1)^2}\bigg(\:}+48 N_s^2 -400 N_s s -176 N_s +512 s +64\bigg) \notag\\
&\quad +\frac{1}{8}\bigg(1152 N_f^2-704 N_f N_s+2384 N_f+275 N_s^2-1664 N_s+1664\bigg)\Bigg] \notag\\
&-16\,u^3\eta N_f\left(\frac{1}{s(s+1)}-\frac{2}{s^2(s+1)^2}\right)\left(2S_{-2}+S_1\right), \label{eq:full-qq}
\end{align}

\begin{align}
\gamma_{q\sigma}^{(3)} = {}
&\dfrac{2u\lambda^2\,N_f (N_s+2)}{s(s+1)}\biggl[ -\frac{2 s+3}{3 s(s+1)} +\frac{4}{3}\biggr] \notag\\
&+8u^2\lambda\,N_f (N_s+2)\biggl[ \frac{2}{3}S_{1} +\frac{2}{3 s(s+1)} -\frac{7}{3}\biggr] \notag\\
&+u^3\,N_f\Bigg[ \bigg(\frac{16 C}{s(s+1)}\bigg)S_{3} +\bigg(\frac{32 C}{s(s+1)}\bigg)\bigg(S_{1,-2}-\tfrac12 S_{-3}\bigg) \notag\\
&\quad -\bigg(\frac{32 C}{s^2(s+1)^2} -\frac{16 \left(N_s^2-2 N_s+8\right)}{s(s+1)}\bigg)S_{-2} \notag\\
&\quad +\bigg((N_s-4) (4 N_f-5 N_s+16)\bigg)S_1^2 \notag\\
&\quad +\bigg(\frac{8 (N_s-4) N_s}{s(s+1)} +(N_s-4) (4 N_f-9 N_s+32)\bigg)S_{2} \notag\\
&\quad +\bigg(\frac{4 (N_s-4) N_s (7 s+4)}{s^2(s+1)^2} -\frac{8 \left(N_s^2-8 N_s+8\right)}{s(s+1)} \notag\\
&\quad \ \ -2 \left(12 N_f N_s-64 N_f-15 N_s^2+120 N_s-192\right)\bigg)S_{1} \notag\\
&\quad -\bigg(\frac{48 C}{s(s+1)} -48 C\bigg)\zeta_3 -\frac{16 N_s^2}{s^4(s+1)^4} \notag\\
&\quad +\frac{1}{s^3(s+1)^3}\bigg(-96 N_f N_s s-80 N_f N_s+96 N_s^2 s+8 N_s^2-96 N_s s-96 N_s\bigg) \notag\\
&\quad +\frac{1}{s^2(s+1)^2}\bigg(152 N_f N_s s +12 N_f N_s +64 N_f -94 N_s^2 s -47 N_s^2 \notag\\
&\quad \phantom{+\frac{1}{s^2(s+1)^2}\bigg(\:} +464 N_s s +184 N_s -256 s  -128\bigg) \notag\\
&\quad -\frac{1}{s(s+1)}\bigg(40 N_f N_s-32 N_f-64 N_s^2+472 N_s-512\bigg) \notag\\
&\quad -\frac{1}{4}\bigg(44 N_f N_s+896 N_f+269 N_s^2-1792 N_s+1792\bigg)\Bigg], \label{eq:full-qs}
\end{align}

\begin{align}
\gamma_{\sigma q}^{(3)} = {}
&-\dfrac{u\lambda^2\,N_s (N_s+2)}{s(s+1)}\biggl[ \frac{1}{9}S_{1} +\frac{2 s+1}{6 s^2(s+1)^2} +\frac{s}{9 s(s+1)} -\frac{13}{36}\biggr] \notag\\
&+\dfrac{2u^2\lambda\,N_s (N_s+2)}{s(s+1)}\biggl[ \frac{2}{3}S_{1} +\frac{2 (2 s+1)}{3 s(s+1)} -\frac{7}{3}\biggr] \notag\\
&+\dfrac{u^3\,N_s}{s(s+1)}\Bigg[ \bigg(\frac{4 C}{s(s+1)}\bigg)S_{3} +\bigg(\frac{8 C}{s(s+1)}\bigg)\bigg(S_{1,-2}-\tfrac12 S_{-3}\bigg) \notag\\
&\quad -\bigg(\frac{8 C}{s^2(s+1)^2} +\frac{1}{s(s+1)}\bigg(16 N_f N_s-32 N_f-8 N_s^2+16 N_s-32\bigg)\bigg)S_{-2} \notag\\
&\quad +\bigg(\frac{N_s (4 N_f-N_s)}{2 s(s+1)} -\frac{1}{2}\bigg(24 N_f^2-26 N_f N_s+56 N_f+7 N_s^2-30 N_s+32\bigg)\bigg)S_1^2 \notag\\
&\quad +\bigg(\frac{N_s (-12 N_f+7 N_s-16)}{2 s(s+1)} \notag\\
&\quad \ \ -\frac{1}{2}\bigg(24 N_f^2-34 N_f N_s+88 N_f+11 N_s^2-54 N_s+64\bigg)\bigg)S_{2} \notag\\
&\quad +\bigg(\frac{1}{s^2(s+1)^2}\bigg(-12 N_f N_s s-8 N_f N_s+11 N_s^2 s+6 N_s^2-32 N_s s-16 N_s\bigg) \notag\\
&\quad \ \ -\frac{1}{s(s+1)}\bigg(24 N_f^2 s -34 N_f N_s s +4 N_f N_s +88 N_f s +32 N_f +11 N_s^2 s -5 N_s^2 -54 N_s s \notag\\
&\quad \phantom{+\frac{1}{s^2(s+1)^2}\bigg(\:} +64 s +16\bigg) \notag\\
&\quad \ \ +2 \left(24 N_f^2-24 N_f N_s+72 N_f+7 N_s^2-42 N_s+48\right)\bigg)S_{1} \notag\\
&\quad -\bigg(\frac{12 C}{s(s+1)} -12 C\bigg)\zeta_3 -\frac{4 N_s^2}{s^4(s+1)^4} \notag\\
&\quad +\frac{1}{s^3(s+1)^3}\bigg(-12 N_f N_s s-20 N_f N_s+3 N_s^2 s-4 N_s^2-24 N_s\bigg) \notag\\
&\quad +\frac{1}{4 s^2(s+1)^2}\bigg(96 N_f^2 s +168 N_f N_s s -44 N_f N_s +320 N_f s +64 N_f -80 N_s^2 s +55 N_s^2 \notag\\
&\quad  \phantom{+\frac{1}{s^2(s+1)^2}\bigg(\:}+224 N_s s -120 N_s -128\bigg) \notag\\
&\quad +\frac{1}{2 s(s+1)}\bigg(96 N_f^2 s -48 N_f^2 -124 N_f N_s s +28 N_f N_s +416 N_f s -16 N_f +47 N_s^2 s \notag\\
&\quad  \phantom{+\frac{1}{s^2(s+1)^2}\bigg(\:} -5 N_s^2 -304 N_s s +12 N_s +384 s\bigg) \notag\\
&\quad +\frac{1}{16}\bigg(44 N_f N_s-1728 N_f-239 N_s^2+1792 N_s-1792\bigg)\Bigg], \label{eq:full-sq}
\end{align}

\begin{align}
\gamma_{\sigma\sigma}^{(3)} = {}
&2\gamma_\sigma^{(3)} \:-\dfrac{\lambda^3\,((N_s+2) (N_s+8))}{s(s+1)}\biggl[ \frac{2}{27}S_{1} +\frac{2 s+1}{18 s(s+1)} -\frac{5}{27}\biggr] \notag\\
&-\dfrac{4u\lambda^2\,N_f (N_s+2)}{s(s+1)}\biggl[ \frac{1}{3}S_{1} +\frac{s}{3 s(s+1)} -1\biggr] \notag\\
&+\dfrac{8u^2\lambda\,N_f (N_s+2)}{s(s+1)}\biggl[ 2S_{1} +\frac{2 s+1}{s(s+1)} -\frac{8}{3}\biggr] \notag\\
&+\dfrac{2u^3\,N_f}{s(s+1)}\Bigg[ 8 CS_{3} +16 C\bigg(S_{1,-2}-\tfrac12 S_{-3}\bigg) -\bigg(\frac{16 C}{s(s+1)} +16 (N_s-2)^2\bigg)S_{-2} \notag\\
&\quad +N_s (4 N_f-N_s)S_1^2 +\bigg(N_s (-12 N_f+7 N_s-16)\bigg)S_{2} \notag\\
&\quad +\bigg(\frac{1}{s(s+1)}\bigg(-24 N_f N_s s-16 N_f N_s+20 N_s^2 s+12 N_s^2-56 N_s s-32 N_s\bigg) \notag\\
&\quad  \phantom{+\frac{1}{s^2(s+1)^2}\bigg(\:}-4 (N_s-2) (4 N_f-N_s+8)\bigg)S_{1} \notag\\
&\quad -24 C\zeta_3 -\frac{8 N_s^2}{s^3(s+1)^3} \notag\\
&\quad +\frac{1}{s^2(s+1)^2}\bigg(24 N_f N_s s+8 N_f N_s+18 N_s^2 s-12 N_s^2-48 N_s s-48 N_s\bigg) \notag\\
&\quad +\frac{1}{s(s+1)}\bigg(48 N_f N_s s+16 N_f N_s-16 N_f-20 N_s^2 s+8 N_s^2+80 N_s s-24 N_s+64 s+32\bigg) \notag\\
&\quad +8 \left(4 N_f N_s-4 N_f-N_s^2+2 N_s-16\right)\Bigg].\label{eq:full-ss}
\end{align}

\if{1==0}
\begin{align}
\gamma_{qq}^{(3)} = {}
&2\gamma_q^{(3)}+u\lambda^2\,N_s (N_s+2)\biggl[ -\frac{2 s+1}{18 s^2 (s+1)^2} +\frac{17}{72 s (s+1)} \biggr] \notag\\
&+2\,u^2\lambda\,N_s (N_s+2)\biggl[ \frac{4}{3 s (s+1)} S_1 +\frac{2 (2 s+1)}{3 s^2 (s+1)^2} -\frac{3}{s (s+1)} \biggr] \notag\\
&-12\,u^3\zeta_3\,C N_s\,\frac{1}{s^2 (s+1)^2} \notag\\
&-8\,u^3\,N_f^2 N_s\biggl[ -2\left(-\frac{2 s+1}{2 s^2 (s+1)^2}-\frac{3}{2 s^2 (s+1)^2}+\frac{4}{s (s+1)}\right)S_1 \notag\\
&\quad +\frac{1}{s (s+1)} S_1^2 +\frac{1}{s (s+1)} S_2 +(2 s+1) \left(\frac{3}{s^3 (s+1)^3}-\frac{7}{s^2 (s+1)^2}\right) \notag\\
&\quad -\frac{3}{s^2 (s+1)^2} +\frac{18}{s (s+1)} \biggr] \notag\\
&+2\,u^3\,N_f N_s\biggl[ -\biggl(-\frac{4 (N_s-3) (2 s+1)}{s^2 (s+1)^2} -\frac{2 (7 N_s-18)}{s^2 (s+1)^2} +\frac{9 (3 N_s-8)}{s (s+1)}\biggr)S_1 \notag\\
&\quad +\frac{N_s}{s (s+1)} S_1^2 +\frac{7 N_s-24}{s (s+1)} S_2 \notag\\
&\quad +(2 s+1) \left(-\frac{12 N_s}{s^4 (s+1)^4}+\frac{41 N_s-40}{s^3 (s+1)^3}+\frac{184-61 N_s}{2 s^2 (s+1)^2}\right) \notag\\
&\quad -\frac{10 N_s}{s^4 (s+1)^4} +\frac{-13 N_s-24}{s^3 (s+1)^3} +\frac{32-45 N_s}{2 s^2 (s+1)^2} +\frac{44 N_s-149}{s (s+1)} \biggr] \notag\\
&-16\,u^3\,\eta N_f\left(\frac{1}{s(s+1)}-\frac{2}{s^2(s+1)^2}\right)\left(2S_{-2}+S_1\right) \notag\\
&-u^3\,N_s\biggl[ -\biggl(\frac{-11 N_s^2+40 N_s-16}{s^2 (s+1)^2} +\frac{41 N_s^2-248 N_s+288}{2 s (s+1)} \notag\\
&\qquad\quad +(2 s+1) \left(\frac{4 (N_s-4) N_s}{s^3 (s+1)^3}-\frac{(N_s-4) (3 N_s-8)}{s^2 (s+1)^2}\right)\biggr)S_1 \notag\\
&\quad +\frac{(N_s-4)^2}{s (s+1)} S_1^2 +\left(N_s-4\right)\left(\frac{3 N_s-8}{s (s+1)}-\frac{2 N_s}{s^2 (s+1)^2}\right)S_2 \notag\\
&\quad +4\left(\frac{2 \left(N_s^2-6 N_s+4\right)}{s^3 (s+1)^3}+\frac{5 N_s^2-26 N_s+16}{s^2 (s+1)^2}\right)S_{-2} \notag\\
&\quad -4 C\left(\frac{1}{s^2 (s+1)^2}+\frac{1}{s (s+1)}\right)S_3 \notag\\
&\quad +4 C\left(\frac{1}{s^2 (s+1)^2}+\frac{1}{s (s+1)}\right)S_{-3} \notag\\
&\quad -8 C\left(\frac{1}{s^2 (s+1)^2}+\frac{1}{s (s+1)}\right)S_{1,-2} \notag\\
&\quad +(2 s+1) \biggl(-\frac{9 N_s^2}{s^4 (s+1)^4} +\frac{-65 N_s^2+400 N_s-512}{4 s^2 (s+1)^2} \notag\\
&\qquad\quad +\frac{(25 N_s-72) N_s}{s^3 (s+1)^3}\biggr) +\frac{4 N_s^2}{s^5 (s+1)^5} +\frac{N_s (5 N_s+24)}{s^4 (s+1)^4} \notag\\
&\quad +\frac{-19 N_s^2+56 N_s+32}{s^3 (s+1)^3} +\frac{-31 N_s^2-48 N_s+384}{4 s^2 (s+1)^2} \notag\\
&\quad +\frac{275 N_s^2-1664 N_s+1664}{8 s (s+1)} \biggr] \label{eq:full-qq}
\end{align}

\begin{align}
\gamma_{q\sigma}^{(3)} = {}
&2\,u\lambda^2\,N_f (N_s+2)\biggl[ -\frac{2}{3 s^2 (s+1)^2} -\frac{2 s+1}{3 s^2 (s+1)^2} +\frac{4}{3 s (s+1)} \biggr] \notag\\
&+8\,u^2\lambda\,N_f (N_s+2)\biggl[ \frac{2}{3} S_1 -\frac{7}{3} +\frac{2}{3 s (s+1)} \biggr] \notag\\
&+48\,u^3\zeta_3\,C N_f\biggl[ 1 -\frac{1}{s (s+1)} \biggr] \notag\\
&+u^3\,N_f^2\biggl[ -8 (3 N_s-16) S_1 +4 (N_s-4) S_1^2 +4 (N_s-4) S_2 -224 -11 N_s \notag\\
&\quad +(2 s+1) \left(\frac{76 N_s}{s^2 (s+1)^2}-\frac{48 N_s}{s^3 (s+1)^3}\right) -\frac{32 N_s}{s^3 (s+1)^3} -\frac{64 (N_s-1)}{s^2 (s+1)^2} \notag\\
&\quad -\frac{8 (5 N_s-4)}{s (s+1)} \biggr] \notag\\
&-u^3\,N_f\biggl[ -2\biggl(-\frac{4 \left(N_s^2-8 N_s+8\right)}{s (s+1)} +3 \left(5 N_s^2-40 N_s+64\right) \notag\\
&\qquad\quad +\frac{7 (N_s-4) N_s (2 s+1)}{s^2 (s+1)^2} +\frac{(N_s-4) N_s}{s^2 (s+1)^2}\biggr)S_1 \notag\\
&\quad +(N_s-4) (5 N_s-16) S_1^2 +\left(N_s-4\right)\left(-\frac{8 N_s}{s (s+1)}+9 N_s-32\right)S_2 \notag\\
&\quad -16\left(\frac{N_s^2-2 N_s+8}{s (s+1)}-\frac{2 \left(N_s^2-6 N_s+4\right)}{s^2 (s+1)^2}\right)S_{-2} -\frac{16 C}{s (s+1)} S_3 \notag\\
&\quad +\frac{16 C}{s (s+1)} S_{-3} -\frac{32 C}{s (s+1)} S_{1,-2} +\frac{1}{4} \left(269 N_s^2-1792 N_s+1792\right) \notag\\
&\quad +(2 s+1) \left(\frac{47 N_s^2-232 N_s+128}{s^2 (s+1)^2}-\frac{48 (N_s-1) N_s}{s^3 (s+1)^3}\right) +\frac{16 N_s^2}{s^4 (s+1)^4} \notag\\
&\quad +\frac{8 N_s (5 N_s+6)}{s^3 (s+1)^3} +\frac{48 N_s}{s^2 (s+1)^2} -\frac{8 \left(8 N_s^2-59 N_s+64\right)}{s (s+1)} \biggr] \label{eq:full-qs}
\end{align}

\begin{align}
\gamma_{\sigma q}^{(3)} = {}
&-u\lambda^2\,N_s (N_s+2)\biggl[ \frac{1}{9 s (s+1)} S_1 +\frac{1}{9 s (s+1)^2} +\frac{2 s+1}{6 s^3 (s+1)^3} \notag\\
&\quad -\frac{13}{36 s (s+1)} \biggr] \notag\\
&+2\,u^2\lambda\,N_s (N_s+2)\biggl[ \frac{2}{3 s (s+1)} S_1 +\frac{2 (2 s+1)}{3 s^2 (s+1)^2} -\frac{7}{3 s (s+1)} \biggr] \notag\\
&+12\,u^3\zeta_3\,C N_s\biggl[ -\frac{1}{s^2 (s+1)^2} +\frac{1}{s (s+1)} \biggr] \notag\\
&-12\,u^3\,N_f^2 N_s\biggl[ -2\left(\frac{2}{s (s+1)}-\frac{1}{s (s+1)^2}\right)S_1 +\frac{1}{s (s+1)} S_1^2 \notag\\
&\quad +\frac{1}{s (s+1)} S_2 -\frac{2}{s^2 (s+1)^3} +\frac{4}{s^2 (s+1)^2} -\frac{2 (2 s+1)}{s^2 (s+1)^2} \biggr] \notag\\
&+u^3\,N_f N_s\biggl[ -2\biggl(\frac{N_s}{s^3 (s+1)^3} +\frac{3 (7 N_s-4)}{2 s^2 (s+1)^2} \notag\\
&\qquad\quad +(2 s+1) \left(\frac{3 N_s}{s^3 (s+1)^3}+\frac{44-17 N_s}{2 s^2 (s+1)^2}\right) +\frac{24 (N_s-3)}{s (s+1)}\biggr)S_1 \notag\\
&\quad +\left(\frac{2 N_s}{s^2 (s+1)^2}+\frac{13 N_s-28}{s (s+1)}\right)S_1^2 +\left(\frac{17 N_s-44}{s (s+1)}-\frac{6 N_s}{s^2 (s+1)^2}\right)S_2 \notag\\
&\quad -\frac{16 (N_s-2)}{s^2 (s+1)^2} S_{-2} \notag\\
&\quad +(2 s+1) \left(-\frac{6 N_s}{s^4 (s+1)^4}+\frac{21 N_s+40}{s^3 (s+1)^3}+\frac{104-31 N_s}{s^2 (s+1)^2}\right) \notag\\
&\quad -\frac{14 N_s}{s^4 (s+1)^4} -\frac{8 (4 N_s+3)}{s^3 (s+1)^3} +\frac{45 N_s-112}{s^2 (s+1)^2} +\frac{11 N_s-432}{4 s (s+1)} \biggr] \notag\\
&-u^3\,N_s\biggl[ -\biggl(\frac{N_s^2}{2 s^3 (s+1)^3} +\frac{21 N_s^2-54 N_s+32}{2 s^2 (s+1)^2} +\frac{2 \left(7 N_s^2-42 N_s+48\right)}{s (s+1)} \notag\\
&\qquad\quad +(2 s+1) \left(\frac{N_s (11 N_s-32)}{2 s^3 (s+1)^3}-\frac{(N_s-2) (11 N_s-32)}{2 s^2 (s+1)^2}\right)\biggr)S_1 \notag\\
&\quad +\left(\frac{N_s^2}{2 s^2 (s+1)^2}+\frac{(N_s-2) (7 N_s-16)}{2 s (s+1)}\right)S_1^2 \notag\\
&\quad +\left(\frac{(N_s-2) (11 N_s-32)}{2 s (s+1)}-\frac{N_s (7 N_s-16)}{2 s^2 (s+1)^2}\right)S_2 \notag\\
&\quad -8\left(\frac{-N_s^2+6 N_s-4}{s^3 (s+1)^3}+\frac{N_s^2-2 N_s+4}{s^2 (s+1)^2}\right)S_{-2} -\frac{4 C}{s^2 (s+1)^2} S_3 \notag\\
&\quad +\frac{4 C}{s^2 (s+1)^2} S_{-3} -\frac{8 C}{s^2 (s+1)^2} S_{1,-2} \notag\\
&\quad +(2 s+1) \biggl(-\frac{3 N_s^2}{2 s^4 (s+1)^4} +\frac{-47 N_s^2+304 N_s-384}{4 s^2 (s+1)^2} \notag\\
&\qquad\quad +\frac{2 (5 N_s-14) N_s}{s^3 (s+1)^3}\biggr) +\frac{4 N_s^2}{s^5 (s+1)^5} +\frac{N_s (11 N_s+48)}{2 s^4 (s+1)^4} \notag\\
&\quad +\frac{-95 N_s^2+232 N_s+128}{4 s^3 (s+1)^3} +\frac{57 N_s^2-328 N_s+384}{4 s^2 (s+1)^2} \notag\\
&\quad +\frac{239 N_s^2-1792 N_s+1792}{16 s (s+1)} \biggr] \label{eq:full-sq}
\end{align}

\begin{align}
\gamma_{\sigma\sigma}^{(3)} = {}
&2\gamma_\sigma^{(3)}+-\lambda^3\,(N_s+2) (N_s+8)\biggl[ \frac{2}{27 s (s+1)} S_1 +\frac{2 s+1}{18 s^2 (s+1)^2} -\frac{5}{27 s (s+1)} \biggr] \notag\\
&-4\,u\lambda^2\,N_f (N_s+2)\biggl[ \frac{1}{3 s (s+1)} S_1 +\frac{1}{3 s (s+1)^2} -\frac{1}{s (s+1)} \biggr] \notag\\
&+8\,u^2\lambda\,N_f (N_s+2)\biggl[ \frac{2}{s (s+1)} S_1 +\frac{2 s+1}{s^2 (s+1)^2} -\frac{8}{3 s (s+1)} \biggr] \notag\\
&-48\,u^3\zeta_3\,C N_f\,\frac{1}{s (s+1)} \notag\\
&+8\,u^3\,N_f^2\biggl[ -2\left(\frac{3 N_s (2 s+1)}{2 s^2 (s+1)^2}+\frac{N_s}{2 s^2 (s+1)^2}+\frac{2 (N_s-2)}{s (s+1)}\right)S_1 \notag\\
&\quad +\frac{N_s}{s (s+1)} S_1^2 -\frac{3 N_s}{s (s+1)} S_2 +(2 s+1) \left(\frac{3 N_s}{s^3 (s+1)^3}+\frac{6 N_s}{s^2 (s+1)^2}\right) \notag\\
&\quad -\frac{N_s}{s^3 (s+1)^3} -\frac{2 (N_s+2)}{s^2 (s+1)^2} +\frac{8 (N_s-1)}{s (s+1)} \biggr] \notag\\
&-2\,u^3\,N_f\biggl[ -4\biggl(\frac{N_s (N_s-2)}{2 s^2 (s+1)^2} +\frac{N_s (5 N_s-14) (2 s+1)}{2 s^2 (s+1)^2} \notag\\
&\qquad\quad +\frac{(N_s-8) (N_s-2)}{s (s+1)}\biggr)S_1 +\frac{N_s^2}{s (s+1)} S_1^2 -\frac{N_s (7 N_s-16)}{s (s+1)} S_2 \notag\\
&\quad +16\left(\frac{N_s^2-6 N_s+4}{s^2 (s+1)^2}+\frac{(N_s-2)^2}{s (s+1)}\right)S_{-2} -\frac{8 C}{s (s+1)} S_3 \notag\\
&\quad +\frac{8 C}{s (s+1)} S_{-3} -\frac{16 C}{s (s+1)} S_{1,-2} \notag\\
&\quad +(2 s+1) \left(\frac{2 \left(5 N_s^2-20 N_s-16\right)}{s^2 (s+1)^2}-\frac{3 N_s (3 N_s-8)}{s^3 (s+1)^3}\right) +\frac{8 N_s^2}{s^4 (s+1)^4} \notag\\
&\quad +\frac{3 N_s (7 N_s+8)}{s^3 (s+1)^3} -\frac{2 N_s (9 N_s-32)}{s^2 (s+1)^2} +\frac{8 \left(N_s^2-2 N_s+16\right)}{s (s+1)} \biggr] \label{eq:full-ss}
\end{align}
\fi


\newpage

\providecommand{\href}[2]{#2}\begingroup\raggedright\endgroup

\end{document}